\documentclass[12pt]{article}
\usepackage{cite}
\usepackage{amsmath, amsthm, amssymb,slashed,url}
\usepackage{ifpdf}
\ifpdf
  \usepackage[pdftex]{graphicx}
  \usepackage{epstopdf}
\else  
  \usepackage[dvips]{graphicx}
\fi
\parskip=\baselineskip

\def\Tr{{\rm Tr}}
\def\16{{\bf 16}}
\def\1{{\bf 1}}
\def\2{{\bf 2}}
\def\3{{\bf 3}}
\def\4{{\bf 4}}

\def\x{{\sf x}}
\def\y{{\sf y}}

\def\a{{\sf a}}
\def\D{{\sf Y}}

\def\tr{{\mathrm{tr}}}

\def\h{\widehat}
\def\bar{\overline}

\def\ra{\rangle}

\def\bp{\begin{pmatrix}}
\def\ep{\end{pmatrix}}
\def\la{\langle}

\font\teneurm=eurm10 \font\seveneurm=eurm7 \font\fiveeurm=eurm5
\newfam\eurmfam
\textfont\eurmfam=\teneurm \scriptfont\eurmfam=\seveneurm
\scriptscriptfont\eurmfam=\fiveeurm

\font\teneusm=eusm10 \font\seveneusm=eusm7 \font\fiveeusm=eusm5
\newfam\eusmfam
\textfont\eusmfam=\teneusm \scriptfont\eusmfam=\seveneusm
\scriptscriptfont\eusmfam=\fiveeusm

\font\tencmmib=cmmib10 \skewchar\tencmmib='177
\font\sevencmmib=cmmib7 \skewchar\sevencmmib='177
\font\fivecmmib=cmmib5 \skewchar\fivecmmib='177
\newfam\cmmibfam
\textfont\cmmibfam=\tencmmib \scriptfont\cmmibfam=\sevencmmib
\scriptscriptfont\cmmibfam=\fivecmmib

\numberwithin{equation}{section}
\def\neg{\negthinspace}
\def\d{\mathrm d}

\def\L{{\mathcal L}}

\def\bar{\overline}
\def\A{{\mathcal A}}
\def\bar{\overline}
\begin{document}
\begin{titlepage}
\begin{flushright}
\end{flushright}
\vskip 1.5in
\begin{center}
{\bf\Large{A Mini-Introduction To Information Theory}}
\vskip
0.5cm {Edward Witten} \vskip 0.05in {\small{ \textit{School of
Natural Sciences, Institute for Advanced Study}\vskip -.4cm
{\textit{Einstein Drive, Princeton, NJ 08540 USA}}}
}
\end{center}
\vskip 0.5in
\baselineskip 16pt
\abstract{This article consists of a very short introduction to classical and quantum information theory.   Basic properties of the classical
Shannon entropy and the quantum von Neumann entropy are described, along with related concepts such as classical and quantum relative entropy,
conditional entropy, and mutual information.  A few more detailed topics are considered in the quantum case. }
\date{May, 2018}
\end{titlepage}
\def\Hom{\mathrm{Hom}}
\def\H{{\mathcal H}}
\def\d{{\mathrm d}}
\def\t{\widetilde}
\def\U{{\mathcal U}}
\def\UU{{\mathrm U}}
\def\V{{\mathcal V}}
\def\st{{\sf t}}
\def\O{{\mathcal O}}
\def\i{{\mathrm i}}
\def\A{{\mathcal A}}
\def\be{\begin{equation}}
\def\ee{\end{equation}}

\tableofcontents

\def\aa{{\sf a}}
\def\bb{{\sf b}}
\def\a{{\bf a}}
\def\x{{\bf x}}
\def\y{{\bf y}}
\def\z{{\bf z}}
 \def\sym{{\mathrm{sym}}}
 \def\zotimes{{\otimes N}}
 \def\cl{{\mathrm{cl}}}
\def\h{\widehat}
\def\t{\widetilde}
\def\CC{{\mathcal C}}
\def\tr{{\mathrm{tr}}}
\def\d{{\mathrm{d}}}
\def\H{{\mathcal H}}
\def\O{{\mathcal O}}
\def\Tr{{\mathrm{Tr}}}
\def\diag{{\mathrm{diag}}}
\def\la{\langle}
\def\ra{\rangle}

\section{Introduction}\label{intro}

This article is intended as a very short introduction to basic aspects of classical and quantum information theory.\footnote{The article is based on a lecture at the 2018 summer program
Prospects in Theoretical Physics at the Institute for Advanced Study.}
 
Section \ref{classic} contains a very short introduction to classical information theory, focusing on the definition of Shannon entropy and related concepts
such as conditional entropy, relative entropy, and mutual information.  Section \ref{quantit} describes the corresponding quantum concepts -- the von Neumann
entropy and the quantum conditional entropy, relative entropy, and mutual information.  Section \ref{more topics} is devoted to some more detailed topics
in the quantum case, chosen to explore the extent to which the quantum concepts match the intuition that their names suggest.

There is much more to say about classical and quantum information theory than can be found here.
There are several excellent introductory  books, for example \cite{NC,Cover, Wilde}.  Another excellent place to start is  the lecture notes \cite{Preskill}, especially
chapter 10.

\section{Classical Information Theory}\label{classic}

\subsection{Shannon Entropy}

We begin with a basic introduction to classical information theory.
 Suppose that one receives a message that consists of a string of symbols $a$ or $b$, say
\begin{equation}\label{One} aababbaaaab \cdots \end{equation}
And let us suppose that $a$ occurs with probability $p$, and $b$ with probability $1-p$.   
How many bits of information can one extract from a long message of this kind, say with $N$
letters?

For large $N$, the message will consist very nearly of $pN$ occurrences of $a$  and $(1-p)N$ occurrences of $b$.      The number
of such messages is 
\begin{align}\frac{N!}{(pN)!((1-p)N)!}&\sim \frac{N^N}{(pN)^{pN}((1-p)N)^{(1-p)N}}\cr &= \frac{1}{p^{pN}(1-p)^{(1-p)N}}=2^{NS}\end{align}
where $S$ is the {\bf Shannon entropy} per letter \cite{Shannon}

\begin{equation}\label{Two} S = -p \log p -(1-p) \log (1-p).\end{equation}
(In information theory, one usually measures entropy in bits and uses logarithms in base 2.)  

The total number of messages of length $N$, given our knowledge of the relative probability of letters $a$ and $b$,
is roughly
\begin{equation}\label{Three} 2^{N S} \end{equation}
and so the number of bits of information one gains in actually
observing such a message is
\begin{equation}\label{Four} NS. \end{equation}
This is an asymptotic formula for large $S$, since we used only the leading term in Stirling's formula to estimate the number of
possible messages, and we ignored fluctuations in the frequencies of the letters.

Suppose more generally that the message is taken from an alphabet with $k$ letters $a_1$, $a_2$, $\cdots$, $a_k$, where
the probability to observe $a_i$ is $p_i$, for $i=1,\cdots, k$.    We write $A$ for this probability distribution.  
   In a long message with $N\gg 1$ letters, the symbol $a_i$ will occur
approximately $Np_i$ times, and the number of such messages is
asymptotically
\begin{equation}\label{Five} \frac{N!}{(p_1N)!(p_2N)!\cdots (p_kN)!} \sim
  \frac{N^N}{\prod_{i=1}^k(p_iN)^{p_iN}}=2^{NS_A} \end{equation}
where now the entropy per letter is 
\begin{equation}\label{Six}  S_A=-\sum_{i=1}^k p_i \log
  p_i. \end{equation}

This is the general definition of the Shannon entropy of a probability distribution for a random variable $A$ that takes values $a_1,\dots,a_k$
with probabilities $p_1,\dots,p_k$.     The number of bits of information that one can extract from a message with $N$ symbols is again
\begin{equation}\label{Seven} NS_A .\end{equation} 
From the derivation, since the number $2^{NS_A}$ of possible messages
is certainly at least 1, we have
\begin{equation}\label{Eight} S_A\geq 0 \end{equation}
for any probability distribution.  To get $S_A=0$, there has to be only 1 possible message, meaning that one of the letters has probability 1
and the others have probability 0.    The maximum possible entropy, for an alphabet with $k$ letters, occurs if the $p_i$ are all $1/k$
and is
\begin{equation}\label{Nine} S_A=-\sum_{i=1}^k (1/k)\log (1/k)= \log
    k. \end{equation}
The reader can prove this by using the method of Lagrange multipliers to
maximize $S_A=-\sum_i p_i\log p_i$ with the constraint $\sum_i p_i=1$.

In engineering applications, $NS_A$ is the number of bits to which a message with $N$ letters can be compressed.     In such applications,
the message is typically not really random but contains information that one wishes to convey.    However, in ``lossless encoding,'' the encoding program does not
understand the message and treats it as random.  
It is easy to imagine a situation in which one can make a better model by incorporating short range correlations between the letters.   (For instance, the
``letters'' might be words in a message in the English language; then English grammar and syntax would dictate short range correlations.  This situation  was actually
considered  by Shannon in his  original paper on this  subject.)  
 A model incorporating such correlations would be a 1-dimensional classical spin chain of some kind with short range interactions.   
Estimating the entropy of a long message of $N$ letters would be a problem in classical statistical mechanics.     But in the ideal gas limit, in which we ignore
correlations, the entropy of a long message is just $NS$ where $S$ is the entropy of a message consisting of only one letter.

Even in the ideal gas model, we are making statements that are only natural in the limit of large $N$.   To formalize the analogy with statistical mechanics,
one could introduce a classical Hamiltonian $H$ whose value for the $i^{th}$ symbol $a_i$ is $-\log p_i$, so that the probability of the $i^{th}$ symbol in the thermodynamic
ensemble is $2^{-H(a_i)}=p_i$.      Notice then that in estimating the number of possible messages for large $N$, we ignored the difference between
the canonical ensemble (defined by probabilities $2^{-H}$) and the microcanonical ensemble (in which one specifies the precise numbers of occurrences of different letters).
    As is usual in statistical mechanics, the different ensembles are equivalent for large $N$.     The equivalence between the different ensembles is important
in classical and quantum information theory.

\subsection{Conditional Entropy}

Now let us consider the following situation.   Alice is trying to communicate with Bob, and she sends a message that consists of many letters,
each being an instance of a random variable\footnote{Generically, a random variable will be denoted $X,Y,Z$, etc.  The probability to observe $X=x$
is denoted $P_X(x)$, so if $x_i$, $i=1,\cdots,n$ are the possible values of $X$, then $\sum_i P_X(x_i)=1$.  Similarly, if $X,Y$ are two random variables,
the probability to observe $X=x$, $Y=y$ will be denoted $P_{X,Y}(x,y)$.} 
$X$ whose possible values are $x_1,\cdots, x_k$.   She sends the message over a noisy telephone connection,
and what Bob receives is many copies of a random variable $Y$, drawn from an alphabet with letters $y_1,\cdots, y_r$.   (Bob might confuse some of Alice's letters
and misunderstand others.)    How many bits of information does Bob gain after Alice has transmitted a message with $N$ letters?
 
To analyze this, let us suppose that $P_{X,Y}(x_i,y_j)$ is the probability that, in a given occurrence, Alice sends $X=x_i$ and Bob hears $Y=y_j$.    The probability
that Bob hears $Y=y_j$, summing over all choices of what Alice
intended, is
\begin{equation}\label{Ten} P_Y(y_j)=\sum_i P_{X,Y}(x_i,y_j). \end{equation}      
If Bob does hear $Y=y_j$, his estimate of the probability that Alice
sent $x_i$ is the {\bf conditional probability}
\begin{equation}\label{Eleven} P_{X|Y}(x_i|y_j)=\frac{P_{X,Y}(x_i,y_j)}{P_Y(y_j)}. \end{equation}
From Bob's point of view, once he has heard $Y=y_j$, his estimate of the remaining entropy in Alice's signal is the Shannon entropy of the conditional
probability distribution.  This is
\begin{equation}\label{Twelve}  S_{X|Y=y_j}=-\sum_i  P_{X|Y}(x_i|y_j)\log
  (P_{X|Y}(x_i|y_j)).\end{equation}

Averaging over all possible values of $Y$, the average remaining entropy, once Bob has heard $Y$, is
\begin{align}\label{number1} \sum_j P_Y(y_j)S_{X|Y=y_j}& =-\sum_j P_Y(y_j) \sum_i \frac{P_{X,Y}(x_i,y_j)}{P_Y(y_j)}\log \left( \frac{P_{X,Y}(x_i,y_j)}{P_Y(y_j)}\right)\cr
&=-\sum_{i,j}P_{X,Y}(x_i,y_j)\log P_{X,Y}(x_i,y_j)+\sum_{i,j} P_{X,Y}(x_i,y_j)\log
P_Y(y_j) \cr &= S_{XY}-S_Y. \end{align}
Here $S_{XY}$ is the entropy of the joint distribution $P_{X,Y}(x_i,y_j)$ for the pair $X,Y$ and $S_Y$ is the entropy of the probability
distribution $P_Y(y_j)=\sum_i P_{X,Y}(x_i,y_j)$ for $Y$ only. 

The left hand side of eqn. (\ref{number1}), which as we see
equals $S_{XY}-S_Y$, is called the {\bf conditional entropy} $S_{X|Y}$ or $S(X|Y)$; it is the entropy that remains in the probability distribution $X$ once $Y$ is known.
Since it was obtained as a sum of ordinary entropies $S_{X|Y=y_j}$
with positive coefficients, it is clearly positive:
\begin{equation}\label{Thirteen}  S_{XY}-S_Y\geq 0. \end{equation}
(The analogous statement 
is {\it not} true quantum mechanically!)  Since $S_X$ is the total information content in Alice's message, and $S_{XY}-S_Y$ is the information content
that Bob still does not have after observing $Y$, it follows that the
information about $X$ that Bob {\it does} gain when he receives $Y$ is
the difference or
\begin{equation}\label{Fourteen} I(X;Y)
    =S_X-S_{XY}+S_Y. \end{equation}
Here $I(X;Y)$ is called the {\bf mutual information} between $X$ and $Y$.     It measures how much we learn about $X$ by observing $Y$.

This interpretation convinces us that $I(X;Y)$ must be nonnegative.     One can prove this directly but instead I want to deduce it from the properties
of one more quantity, the relative entropy.  This will  complete our cast of characters.

\subsection{Relative Entropy}\label{cre}

One can motivate the definition of relative entropy as follows.   Suppose that we are observing a random variable
$X$, for example the final state in the decays of a radioactive nucleus.      We have a theory that predicts a probability distribution $Q_X$ for the final state,
say the prediction is that the probability to observe final state $X=x_i$, where $i$ runs over a set of possible outcomes $\{1,2,\cdots s\}$,  is $q_i=Q_X(x_i)$.     But maybe our theory is wrong and the decay is actually described by some
different probability distribution $P_X$, such that the probability of $X=x_i$ is $p_i=P_X(x_i)$.     After observing the decays of $N$ atoms, how sure
could we be that the initial hypothesis is wrong? 

If the correct probability distribution is $P_X$, then after observing $N$ decays, we will see outcome $x_i$ approximately $p_iN$ times.  Believing $Q_X$ to
be the correct distribution, we will
judge the probability of what we have seen to be\footnote{Here $\frac{N!}{\prod_{j=1}^s(p_jN)!}$ is the number of sequences in which outcome $x_i$ occurs
$p_iN$ times, and $\prod_{i=1}^s q_i^{p_iN}$ is the probability of any specific such sequence, assuming that the initial hypothesis $Q_X$ is correct.}
\begin{equation}\label{Fifteen} {\mathcal P}=\prod_{i=1}^s
  q_i^{p_iN}\frac{N!}{\prod_{j=1}^s(p_jN)!}. \end{equation}
We already calculated that for large $N$ 
\begin{equation}\label{Sixteen} \frac{N!}{\prod_{j=1}^s(p_jN)!}\sim
  2^{-N\sum_i p_i\log p_i} \end{equation}
so 
\begin{equation}\label{Seventeen} {\mathcal P}\sim 2^{-N\sum_i
    p_i(\log p_i -\log q_i)}. \end{equation}
This is $2^{-N S(P||Q)}$
where the relative entropy (per observation) or Kullback-Liebler
divergence is defined as
\begin{equation}\label{Eightteen} S(P_X||Q_X)=\sum_i p_i (\log p_i - \log q_i). \end{equation}
From the derivation, $S(P_X||Q_X)$ is clearly nonnegative, and zero only if $P_X=Q_X$, that is 
if the initial hypothesis is correct.   If the initial hypothesis is wrong, we will be sure
of this once
\begin{equation}\label{Nineteen} NS(P_X||Q_X)\gg 1.  \end{equation}

The chance of falsely excluding a correct hypothesis, because of a large fluctuation that causes the data to be more accurately
simulated by $P_X$ than by $Q_X$,  decays for large $N$ as $2^{-N S(P_X||Q_X)}$.   (Later we will more loosely say that the confidence in excluding the wrong
hypothesis is controlled by $2^{-N S(P_X||Q_X)}$.) 
In this analysis, we have ignored noise in the observations.  What we learned earlier about conditional entropy would give us a start in including the
effects of noise.

$S(P_X||Q_X)$ is an important measure of the difference between two probability distributions $P_X$ and $Q_X$, but notice that it is asymmetric in $P_X$ and $Q_X$.
We broke the symmetry by assuming that $Q_X$ was our initial hypothesis and $P_X$ was the correct answer.  

Now we will use positivity of the relative entropy to prove positivity of the mutual information.   We consider a pair of random variables $X$, $Y$ and we consider
two different probability distributions.    One, which we will call
$P_{X,Y}$, is defined by a possibly correlated joint probability
distribution
\begin{equation}\label{Twenty} P_{X,Y}(x_i,y_j). \end{equation}
Given such a joint probability distribution, the separate probability distributions for $X$ and for $Y$
are obtained by ``integrating out'' or summing over the other variable:
\begin{equation}\label{Twenty1}P_X(x_i)=\sum_j P_{X,Y}(x_i,y_j),~~~~~
  P_Y(y_j)=\sum_i P_{X,Y}(x_i,y_j). \end{equation}
  This is an important operation which will frequently recur.
We define a second probability distribution for $X,Y$ by ignoring the correlations
between them:
\begin{equation}\label{Twenty2}
  Q_{X,Y}(x_i,y_j)=P_X(x_i)P_Y(y_j). \end{equation}

Now we calculate the relative entropy between these two distributions:
\begin{align}\label{Number3} S(P_{X,Y}||Q_{X,Y})=&\sum_{i,j}P_{X,Y}(x_i,y_j)(\log P_{X,Y}(x_i,y_j) -\log(P_X(x_i) P_Y(y_j)))\cr=&\sum_{i,j}P_{X,Y}(x_i,y_j)(\log P_{X,Y}(x_i,y_j) -\log P_X(x_i)-\log  P_Y(y_j))\cr=
&S_X+S_Y-S_{XY}=I(X;Y).\end{align}
Thus $I(X;Y)\geq 0$, with equality only if the two distributions are the same, meaning that $X$ and $Y$ were uncorrelated to begin with.      

The property
\begin{equation}\label{Twenty3} S_X+S_Y-S_{XY}\geq 0 \end{equation}
is called {\bf subadditivity} of entropy.

\subsection{Monotonicity of Relative Entropy}

Now there is one more very important property of relative entropy that I want to explain, and this will more or less conclude our introduction to classical
information theory.     Suppose that $X$ and $Y$ are two random variables.  Let $P_{X,Y}$ and $Q_{X,Y}$ be two probability distributions,
described by functions $P_{X,Y}(x_i,y_j)$ and $Q_{X,Y}(x_i,y_j)$.     
If we start with a hypothesis $Q_{X,Y}$ for the joint probability, then after many trials in which we observe $X$ and $Y$, our confidence that we are wrong (assuming
that $P_{X,Y}$ is the correct answer) is determined by
$S(P_{X,Y}||Q_{X,Y})$.      But suppose that we only observe $X$ and not $Y$.   The reduced distributions $P_X$ and $Q_X$ for $X$ only are
described by functions
\begin{equation}\label{Twenty4} P_X(x_i)=\sum_j P_{X,Y}(x_i,y_j), ~~~~~
  Q_X(x_i)=\sum_j Q_{X,Y}(x_i,y_j). \end{equation}
If we observe $X$ only, then the confidence after many trials that the initial hypothesis is wrong is controlled by $S(P_X||Q_X)$.

It is harder to disprove the initial hypothesis if we observe only
$X$, so 
\begin{equation}\label{Twenty5} S(P_{X,Y}||Q_{X,Y})\geq S(P_X||Q_X). \end{equation}     
This is called {\bf monotonicity of relative entropy}.

Concretely, if we observe a sequence $x_{i_1},x_{i_2},\dots x_{i_N}$ in $N$ trials, then to estimate how unlikely this is, we will imagine a sequence of $y$'s that minimizes
the unlikelihood of the joint sequence \be\label{pop}(x_{i_1},y_{i_1}),(x_{i_2},y_{i_2}),\cdots 
,(x_{i_N},y_{i_N}).\ee   An actual sequence of $y$'s that we might observe can only
be more unlikely than this.  So observing $Y$ as well as $X$ can only increase our estimate of how unlikely the outcome was, given the sequence of the $x$'s.
Thus, the relative entropy only goes down upon ``integrating out'' some variables and not observing them. 

Hopefully, the reader has found this explanation compelling, but it is also not difficult to give a proof in formulas.
The inequality $S(P_{X,Y}||Q_{X,Y})- S(P_X||Q_X)\geq 0$ can be written
\begin{equation}\label{Twenty6}
  \sum_{i,j}P_{X,Y}(x_i,y_j)\left(\log\left(\frac{
        P_{X,Y}(x_i,y_j)}{Q_{X,Y}(x_i,y_j)}\right)-\log
    \left(\frac{P_X(x_i)}{Q_X(x_i)}\right)   \right)\geq 0. \end{equation}
Equivalently 
\begin{equation}\label{Twenty7} \sum_i P_X(x_i)
  \sum_j\frac{P_{X,Y}(x_i,y_j)}{P_X(x_i)}\log
  \left(\frac{P_{X,Y}(x_i,y_j)/P_X(x_i)}{Q_{X,Y}(x_i,y_j)/Q_X(x_i)}  \right)\geq
  0. \end{equation}
The left hand side is a sum of positive terms, since it is
\begin{equation}\label{Twenty8} \sum_i P_X(x_i)
  S(P_{Y|X=x_i}||Q_{Y|{X=x_i}}), \end{equation}
where we define probability distributions $P_{Y|{X=x_i}}$, $Q_{Y|{X=x_i}}$
 conditional on observing $X=x_i$:
\begin{equation}\label{Twenty9}
  P_{Y|X=x_i}(y_j)={P_{X,Y}(x_i,y_j)}/{P_X(x_i)},~~~
  Q_{Y|X=x_i}(y_j)={Q_{X,Y}(x_i,y_j)}/{Q_X(x_i)}. \end{equation}

So this establishes monotonicity of relative entropy.\footnote{What we have described 
is not the most general statement of monotonicity of relative entropy in classical information
theory.   More generally, relative entropy is monotonic under an arbitrary stochastic map.   We will not explain this here, though later we will explain
the quantum analog (quantum relative entropy is monotonic in any quantum channel).}  
  An important special case is {\bf strong subadditivity} of entropy.   For this, we consider three random variables $X,Y,Z$.   The combined system
has a joint probability distribution $P_{X,Y,Z}(x_i,y_j,z_k)$.  
Alternatively, we could forget the correlations between $X$ and $YZ$,
defining a probability distribution $Q_{X,Y,Z}$ for the system $XYZ$ by
\begin{equation}\label{Thirty}
  Q_{X,Y,Z}(x_i,y_j,z_k)=P_X(x_i)P_{Y,Z}(y_j,z_k) \end{equation}
where  as usual
\begin{equation}\label{Thirty1}  P_X(x_i)=\sum_{j,k}P_{X,Y,Z}(x_i,y_j,z_k),~~~~
  P_{Y,Z}(y_j,z_k)=\sum_i P_{X,Y,Z}(x_i,y_j,z_k). \end{equation}
The relative entropy is  $S(P_{X,Y,Z}||Q_{X,Y,Z})$.  
But what if we only observe the subsystem $XY$?   Then we replace $P_{X,Y,Z}$ and $Q_{X,Y,Z}$ by probability distributions
$P_{X,Y}$, $Q_{X,Y}$ with
\begin{equation}\label{Thirty2} P_{X,Y}(x_i,y_j)=\sum_k P_{X,Y,Z}(x_i,y_j,z_k),
  ~~~~~~ Q_{X,Y}(x_i,y_j)=\sum_k Q_{X,Y,Z}(x_i,y_j,z_k) =P_X(x_i)P_Y(y_j) \end{equation}
and we can define the relative entropy $S(P_{X,Y}||Q_{X,Y})$.
Monotonicity of relative entropy tells us that
\begin{equation}\label{Thirty3} S(P_{X,Y,Z}||Q_{X,Y,Z})\geq
  S(P_{X,Y}||Q_{X,Y}).  \end{equation}

But the relation between relative entropy and mutual information that
we discussed a moment ago gives
\begin{equation}\label{Thirty4}
  S(P_{X,Y,Z}||Q_{X,Y,Z})=I(X;YZ)=S_X-S_{XYZ}+S_{YZ} \end{equation}
and
\begin{equation}\label{Thirty5}
  S(P_{X,Y}||Q_{X,Y})=I(X;Y)=S_X-S_{XY}+S_Y. \end{equation}

So
\begin{equation}\label{Thirty6} S_X-S_{XYZ}+S_{YZ}\geq
  S_X-S_{XY}+S_Y \end{equation}
or 
\begin{equation}\label{Thirty7} S_{XY}+S_{YZ}\geq
  S_Y+S_{XYZ}, \end{equation}
which is called {\bf strong subadditivity}.   Remarkably, the same statement turns out to be true in quantum mechanics, where it is both powerful and surprising.

Equivalently, the comparison of eqns. (\ref{Thirty4}) and (\ref{Thirty5}) gives
\be\label{modo} I(X;YZ)\geq I(X;Y), \ee
which is called monotonicity of mutual information.   The intuition is that what one learns about a random variable $X$ by observing both $Y$ and $Z$
is at least as much as one could learn by observing $Y$ only.

We conclude this mini-introduction to classical information theory with one last remark.     We repeatedly made use of the ability to define a conditional
probability distribution, conditional on some observation.    This has no really close analog in  the quantum mechanical case\footnote{See, however,
\cite{Subs} for a partial substitute.}  and it is something
of a miracle that many of the conclusions nonetheless have quantum mechanical analogs.     The greatest miracle is strong subadditivity of quantum entropy.

\section{Quantum Information Theory: Basic Ingredients}\label{quantit}

\subsection{Density Matrices}

Now we turn to quantum information theory.     Quantum mechanics always deals with probabilities, but the real quantum analog of a classical
probability distribution is not a quantum state but
a {\it  density matrix}.      Depending on one's view of quantum mechanics, one might believe that the whole universe is described
by a quantum mechanical pure state that depends on all the available degrees of freedom.   Even if this is true, one usually studies a subsystem that cannot
be described by a pure state.   

For an idealized case, let $A$ be a subsystem of interest, with Hilbert space $\H_A$.  And let $B$ be everything else of relevance,
or possibly all of the rest of the universe, with Hilbert space $\H_B$.   The combined Hilbert space is the tensor product $\H_{AB}=\H_A\otimes \H_B$. 
The simple case is that a state vector $\psi_{AB}$ of the combined system is the tensor product of a state vector $\psi_A\in\H_A$ and another state vector
$\psi_B\in\H_B$:
\begin{equation}\label{Thirty8} \psi_{AB}=\psi_A\otimes
  \psi_B. \end{equation}   
If $\psi_{AB}$ is a unit vector, we can choose $\psi_A$ and $\psi_B$ to also be unit vectors.
In the case of such a product state, predictions about the $A$ system can be made by forgetting about the $B$ system and using the state vector $\psi_A$.
Indeed, if $\O_A$ is any operator on $\H_A$, then the corresponding operator on $\H_{AB}$ is $\O_A\otimes 1_B$, and its expectation value in a factorized state
$\psi_{AB}=\psi_A\otimes \psi_B$ is
\begin{equation}\label{Thirty9} \la\psi_{AB}|\O_A\otimes 1_B|\psi_{AB}\ra = \la \psi_A|\O_A|\psi_A\ra\la \psi_B|1_B|\psi_B\ra = \la \psi_A|\O_A|\psi_A\ra. \end{equation}

However, a generic pure state $\psi_{AB}\in\H_{AB}$ is not a product state; instead it is ``entangled.''    
If $\H_A$ and $\H_B$ have dimensions $N$ and $M$, then a generic state
in $\H_{AB}$ can be presented as an $N\times M$ matrix, for example in the $2\times 3$
case
\begin{equation}\label{Fourty}  \psi_{AB}= \begin{pmatrix} *&*&*\cr
    *&*&* \end{pmatrix} . \end{equation}
By unitary transformations on $\H_A$ and on $\H_B$, we can transform
$\psi_{AB}$ to
\begin{equation}\label{Fourty1}  \psi_{AB}\to U\psi_{AB}
  V \end{equation}
where $U$ and $V$ are $N\times N$ and $M\times M$ unitaries.  The canonical form of a matrix under that operation is a diagonal matrix, with positive
numbers on the diagonal, and extra rows or columns of zeroes, for example
$$\begin{pmatrix} \sqrt{p_1} & 0 & 0\cr0 & \sqrt{p_2}& 0 \end{pmatrix}.$$

A slightly more invariant way to say this is that any  pure state can
be written 
\begin{equation}\label{Fourty2}
\psi_{AB}=\sum_i \sqrt{p_i}\psi_A^i\otimes \psi_B^i,  \end{equation}
where we can assume that $\psi_A^i$ and $\psi_B^i$ are orthonormal,
\begin{equation}\label{Fourty3} \la \psi_A^i,\psi_A^j\ra
  =\la\psi_B^i,\psi_B^j\ra =\delta^{ij} \end{equation}
and that $p_i>0$.  (The $\psi_A^i$ and $\psi_B^i$ may not be bases of $\H_A$ or $\H_B$, because there may not be enough of them.)  The condition for
$\psi_{AB}$ to be a unit vector is that
\begin{equation}\label{Fourty4}  \sum_i p_i=1, \end{equation}
so we can think of the $p_i$ as probabilities.  Eqn. (\ref{Fourty2}) is called the Schmidt decomposition.

What is the expectation value in such a state of an operator $\O_A$ that only acts on $A$?  It is
\begin{align}\label{number3} \la\psi_{AB}|\O_A\otimes 1_B|\psi_{AB}\ra&=\sum_{i,j} \sqrt{p_ip_j}\la \psi_A^i|\O_A|\psi_A^j\ra \la \psi_B^i|1_B|\psi_B^j\ra \cr&=\sum_ip_i\la\psi_A^i|\O_A|\psi_A^i\ra.
\end{align}
This is the same as
\begin{equation}\label{Fourty5} \Tr_{\H_A}\,\rho_A
  \O_A, \end{equation}
where $\rho_A$ is the {\bf density matrix}
\begin{equation}\label{Fourty6}  \rho_A=\sum_i p_i |\psi_A^i\ra
  \la\psi_A^i|. \end{equation}
Thus, if we are only going to make measurements on system $A$, we do not need a wavefunction of the universe: it is sufficient to have a density matrix
for system $A$.

From the definition
\begin{equation}\label{Fourty7} \rho_A=\sum_i p_i |\psi_A^i\ra
  \la\psi_A^i|\end{equation}
we see that $\rho_A$   is hermitian and positive semi-definite.
Because $\sum_ip_i=1$, $\rho_A$ has trace 1:
\begin{equation}\label{Fourty8} \Tr_{\H_A} \,\rho_A=1. \end{equation}  
Conversely, every matrix with those properties can be ``purified,'' meaning that it is the density matrix of some pure state
on some ``bipartite'' (or two-part) system $AB$.    For this, we first
observe that any hermitian matrix $\rho_A$ can be diagonalized,
meaning that in a suitable basis it takes the
form of eqn. (\ref{Fourty7}); moreover, if $\rho_A\geq 0$, then the $p_i$ are likewise positive (if one of the $p_i$ vanishes, we omit it from the sum).      Having gotten this
far, to realize $\rho_A$ as a density matrix we simply introduce another Hilbert space $\H_B$ with orthonormal states $\psi_B^i$ and observe that $\rho_A$ is the density
matrix of the pure state
\begin{equation}\label{Fourty9} \psi_{AB}=\sum_i\sqrt{p_i}\psi_A^i\otimes \psi_B^i\in \H_A\otimes \H_B. \end{equation}
In this situation, $\psi_{AB}$ is called a ``purification'' of the density matrix $\rho_A$.     The existence of purifications is a nice property of quantum mechanics
that has no classical analog: the classical analog of a density matrix is a probability distribution, and there is no notion of purifying a probability distribution. 

The purification $\psi_{AB}$ of a density matrix $\rho_A$ is far from unique (even if the auxiliary system $B$ is specified), because there is freedom
in choosing the  orthonormal states
$\psi_B^i$ in eqn. (\ref{Fourty9}).   However, any other set of orthonormal vectors in $\H_B$ can be obtained from a given choice $\psi_B^i$ by a unitary
transformation of $\H_B$, so we learn the following important fact: any two purifications of the same density matrix $\rho_A$ on system $A$ by pure states
of a bipartite system $AB$ are equivalent under a unitary transformation of system $B$.
 
If there is more than one term in the expansion 
\begin{equation}\label{Fifty}
  \psi_{AB}=\sum_i\sqrt{p_i}\psi_A^i\otimes \psi_B^i\in \H_A\otimes
  \H_B, \end{equation}
we say that systems $A$ and $B$ are entangled in the state
$\psi_{AB}$.  If there is only one term, the expansion reduces to
\begin{equation}\label{Fifty1} \psi_{AB}=\psi_A\otimes
  \psi_B,  \end{equation}
an ``unentangled'' tensor product state.    Then system $A$ can be described by the pure state $\psi_A$ and the density matrix is of rank 1:
$$\rho_A=|\psi_A\ra\la\psi_A|.$$   If $\rho_A$ has rank higher than 1, we say that system $A$ is in a mixed state.  If $\rho_A$ is a multiple of the identity, we say that $A$ is maximally mixed.

In the general case
\begin{equation}\label{Fifty2} \rho_A=\sum_i p_i
  |\psi_A^i\ra\la\psi_A^i| \end{equation}
one will describe all measurements of system $A$ correctly if one says that system $A$ is in the state $\psi_A^i$ with probability $p_i$.
However, one has to be careful here because the decomposition
of eqn. (\ref{Fifty2})
is not unique.  It is unique if the $p_i$ are all distinct and one wants the number of terms in the expansion to be as small as possible, or equivalently if
one wants the $\psi_A^i$ to be orthonormal.  But if one relaxes those conditions, then (except for a pure state) there are many ways to make this
expansion.    This means that if Alice prepares a quantum system to be in the pure state $\psi_A^i$ with probability $p_i$, then there is no way to determine the $p_i$ or the $\psi_A^i$ by measurements, even if one is provided with many identical copies to measure.   Any
measurement of the system will depend only on $\rho_A=\sum_i p_i
  |\psi_A^i\ra\la\psi_A^i|$.  There is no way to get additional information about how the system was prepared.

So far, when we have discussed a bipartite system $AB$, we have assumed that the combined system is in a pure state $\psi_{AB}$,
and we have discussed density matrices $\rho_A$ and $\rho_B$ for systems $A$ and $B$.   More generally, we should allow for
the possibility that the combined system $AB$ is described to begin with by a density matrix $\rho_{AB}$.  Consideration
of this situation leads to the following very fundamental definition.

Just as for classical probability distributions, for density matrices we can always ``integrate out'' an unobserved system and get a
reduced density matrix for a subsystem.    Classically, given a joint probability distribution $P_{X,Y}(x_i,y_j)$ for a bipartite system $XY$,
we ``integrated out'' $Y$ to get a probability distribution for $X$
only:
\begin{equation}\label{Sixty5}  P_X(x_i)=\sum_j
  P_{X,Y}(x_i,y_j).   \end{equation}
The quantum analog of that is a partial trace.   Suppose that $AB$ is a bipartite system with Hilbert space $\H_A\otimes \H_B$ and a density
matrix $\rho_{AB}$. 
Concretely, if $|i\ra\neg_A$, $i=1,\dots,n$  are an orthonormal  basis of $\H_A$ and $|\alpha\ra\neg_B$, $\alpha=1,\dots, m$
 are an orthonormal basis of $\H_B$, then a density
matrix for $AB$ takes the general form
\begin{equation}\label{Sixty6} 
\rho_{AB}=\sum_{i,i',\alpha,\alpha'} c_{ii'\alpha\alpha'} |i\ra\neg_A
\otimes |\alpha\ra\neg_B ~_A\la {i'}|\otimes\,\neg_B \la
{\alpha'}|.  \end{equation}
The reduced density matrix for measurements of system $A$ only is
obtained by setting $\alpha=\alpha'$, replacing $|\alpha\ra\neg_B\, _B\la\alpha|$ by its trace, which is 1,
 and summing:
\begin{equation}\label{Sixty7} 
\rho_A=\sum_{i,i',\alpha} c_{i,i',\alpha,\alpha} |i\ra\neg_A\,_A\la
{i'}|.   \end{equation}
In other words, if we are going to measure system $A$ only, we sum over
all of the unobserved states of system $B$.
This is usually written as a partial trace:
\begin{equation}\label{Sixty8} 
\rho_A=\Tr_{\H_B} \,\rho_{AB}, 
\end{equation}
the idea being that one has ``traced out'' $\H_B$, leaving a density
operator on $\H_A$.   Likewise (summing over $i$ to eliminate $\H_A$)
\begin{equation}\label{Sixty9} 
\rho_B=\Tr_{\H_A}\,\rho_{AB}. \end{equation}

Before going on, perhaps I should give a simple example of a concrete situation in which it is impractical to not use density matrices.      Consider
an isolated atom interacting with passing photons.  A photon  might be scattered, or absorbed and reemitted, or might pass by without interacting with the atom.   Regardless, 
  after a certain time, the atom is again alone.  After  $n$ photons have had the chance to interact with the atom, to give a pure state description, we need a joint wavefunction for the atom and all the outgoing photons.  The mathematical machinery gets
bigger and bigger, even though (assuming we observe only  the atom) the physical situation is not changing.   By using a density
matrix, we get a mathematical framework for describing the state of the system that does not change regardless of how many photons have interacted with the atom in the past (and what
else those photons might have interacted with).   All we need is a density matrix for the atom.

\subsection{Quantum Entropy}

The {\bf von Neumann entropy}\footnote{The von Neumann entropy is the most important quantum entropy, but generalizations such as the R\'{e}nyi entropies  
$S_\alpha(\rho_A)=\frac{1}{1-\alpha}\log \Tr\, \rho_A^\alpha$
can also be useful.}
of a density matrix $\rho_A$ is defined by a
formula analogous to the Shannon entropy of a probability
distribution:
\begin{equation}\label{Fifty4} S(\rho_A)=-\Tr\,\rho_A\log
  \rho_A.   \end{equation}
  As an immediate comment, we note that $S(\rho_A)$ is manifestly invariant under a unitary transformation
  \be\label{zold}\rho_A\to U\rho_A U^{-1}.\ee
  Quantum conditional and relative entropy, which will be introduced
  in section \ref{crq}, are similarly invariant under a suitable class of unitaries.
  
By a unitary transformation, we can diagonalize $\rho_A$, putting it in the form
\begin{equation}\label{Fifty5} \rho_A=\sum_i p_i
  |\psi_A^i\ra\la\psi_A^i|,  \end{equation}
with $\psi_A^i$ being orthonormal and $p_i>0$.   Then in an obvious basis
\begin{equation}\label{Fifty6}  \rho_A\log \rho_A =\begin{pmatrix} p_1\log p_1 & &&& \cr
                                                                & p_2\log p_2 &&& \cr
                                                                && p_3\log p_3 &&\cr
                                                                  &&&\ddots \end{pmatrix} \end{equation}                                                                
 and             so
\begin{equation}\label{Fifty7} 
S(\rho_A)=-\sum_i p_i \log p_i,  \end{equation}
the same as the Shannon entropy of the probability distribution $\{p_i\}$.  

An immediate consequence is that, just as for the Shannon entropy,
\begin{equation}\label{Fifty8} S(\rho_A)\geq 0, \end{equation} 
with equality only for a pure state (one of the $p$'s  being 1 and the others 0).  
The formula $S(\rho_A)=-\sum_ip_i\log p_i$ also implies the same upper
bound that we had classically for a system with $k$ states
\begin{equation}\label{Fifty9}  S(\rho_A)\leq \log k,  \end{equation}
with equality only if $\rho_A$ is a multiple of the identity:
\begin{equation}\label{Sixty} 
\rho_A=\frac{1}{k}\begin{pmatrix} 1&&&\cr
                                                     & 1 &&\cr && 1&
                                                     \cr
                                                     &&&\ddots\end{pmatrix}.   \end{equation}
In this case, we say that  $A$ is in a maximally mixed state.                                                       
 In fact, the von Neumann entropy has many properties analogous
to the Shannon entropy, but the explanations required are usually more subtle and there are key differences.  

Here is a nice property of the von Neumann entropy that does {\it not} have a classical analog.  If a bipartite system $AB$ is in a pure
state
\begin{equation}\label{Sixty1}
  \psi_{AB}=\sum_i\sqrt{p_i}\psi_A^i\otimes \psi_B^i\in \H_A\otimes
  \H_B,\end{equation}
then the density matrices of systems $A$ and $B$ are
\begin{equation}\label{Sixty2} \rho_A=\sum_i p_i
  |\psi_A^i\ra\la\psi_A^i|,  \end{equation}
and likewise
\begin{equation}\label{Sixty3}  \rho_B=\sum_i p_i
  |\psi_B^i\ra\la\psi_B^i|.  \end{equation}

The same constants $p_i$ appear in each, so clearly
\begin{equation}\label{Sixty4}  S(\rho_A)=S(\rho_B).  \end{equation}
Thus a system $A$ and a purifying system $B$ always have the same entropy.   Note that in this situation, since the combined system $AB$ is in a pure
state, its entropy $S_{AB}$ vanishes.

\subsection{Concavity}\label{convexty}

The von Neumann entropy -- like its antecedents in classical thermodynamics and statistical mechanics -- has the important property of {\bf concavity}. 
Suppose that $\rho_1$ and $\rho_2$ are two density matrices,
and set  $\rho(t)=t\rho_1+(1-t) \rho_2$, for $0\leq t\leq 1$.  We will write $\dot\rho(t)$, $\ddot\rho(t)$ for $\d \rho(t)/\d t$, $\d^2\rho(t)/\d t^2$.
Then
 \be\label{convext}\frac{\d^2}{\d t^2}S(\rho(t))\leq 0.\ee
 To prove this, we first compute that\footnote{\label{similar} For this, consider an arbitrary density matrix $\rho$ and a first order perturbation $\rho\to\rho+\delta\rho$.  After
 diagonalizing $\rho$, one observes that to first order in $\delta\rho$, the off-diagonal part of $\delta\rho$ does not contribute to the trace in the definition
 of $S(\rho+\delta\rho)$.    Therefore, $S(\rho(t))$ can be differentiated assuming that $\rho$ and $\dot\rho$ commute.  So it suffices to check
 (\ref{onvexity})  for a diagonal family of density matrices $\rho(t)=\mathrm{diag}(\lambda_1(t),\lambda_2(t),\cdots,\lambda_n(t))$, with $\sum_i \lambda_i(t)=1$.
 Another approach is to use (\ref{nvexity}) to substitute for $\log \rho(t)$ in the definition $S(\rho(t))=-\Tr\,\rho(t)\log \rho(t)$.   Differentiating with
 respect to $t$, observing that $\rho(t)$ commutes with $1/(s+\rho(t))$,  and then integrating over $s$, one arrives at (\ref{onvexity}).  In either approach, one
 uses that $\Tr\,\dot\rho=0$ since $\Tr\,\rho(t)=1$.}
 \be\label{onvexity} \frac{\d}{\d t}S(\rho(t))=-\Tr\,\dot \rho \log \rho. \ee
 Then as
 \be\label{nvexity}\log\rho =\int_0^\infty\d s\left(\frac{1}{s+1}-\frac{1}{s+\rho(t)}\right)\ee
 and $\ddot \rho=0$, we have
 \be\label{vexity}\frac{\d^2}{\d t^2}S(\rho(t)) =-\int_0^\infty \d s\Tr\, \dot \rho \frac{1}{s+\rho(t)} \dot \rho \frac{1}{s+\rho(t)}.\ee
 The integrand is positive, as it is $\Tr\,B^2$, where $B$ is the  self-adjoint operator $(s+\rho(t))^{-1/2}\dot\rho(t) (s+\rho(t))^{-1/2}$.
 So
 $\frac{\d^2}{\d t^2} S(\rho(t)) \leq 0.$

 In other words, the function $S(\rho(t))$ is concave.  Like any concave function, $S(\rho(t))$ has the property that the straight line connecting
 two points on its graph lies below the graph.  Explicitly, this gives
 \be\label{xity} t S(\rho_1)+(1-t)S(\rho_2) \leq S(t\rho_1+(1-t)\rho_2)= S(\rho(t)). \ee
 More generally, let $\rho_i$, $i=1,\dots, n$ be density matrices and $p_i$, $i=1,\dots, n$ nonnegative numbers with $\sum_i p_i=1$.
 Then by induction starting with (\ref{xity}), or because this is a general property of concave functions, we have
 \be\label{with}\sum_i p_i S(\rho_i)\leq S(\rho),~~~\rho=\sum_i p_i\rho_i. \ee
 This may be described by saying that entropy can only increase under mixing. 
The nonnegative quantity that appears here  is known as the {\bf Holevo information} or
Holevo $\chi$ \cite{Holevo}:
\begin{equation}\label{holevo}\chi=S(\rho)-\sum_i p_i S(\rho_i). \ee
 
An interesting special case is the following.   Let $\rho$ be any density matrix on a Hilbert space $\H$.  Pick a basis of $\H$, and let $\rho_D$ be
the diagonal density matrix obtained in that basis by dropping the off-diagonal matrix elements from $\rho$ and keeping the diagonal ones.
Let $\rho(t)=(1-t)\rho_D+t\rho.$   We see that
\be\label{wort}\left.\frac{\d}{\d t}S(\rho(t)) \right|_{t=0}=0, \ee
by virtue of (\ref{onvexity}), 
because $\rho(0)$  and  $\log \rho(0)$ are diagonal while the diagonal matrix elements of $\d\rho/\d t$ vanish at $t=0$.
When we combine this with $\d^2 S(\rho(t))/\d t^2\leq 0$, we get $S(\rho(1))\leq S(\rho(0))$ or
\be\label{nort} S(\rho_D)\geq S(\rho). \ee
Thus, dropping the off-diagonal part of a density matrix (in any basis) can only increase the entropy.   Eqn. (\ref{nort}) is a strict inequality unless $\rho=\rho_D$,
because eqn. (\ref{vexity}) shows that $\left.\frac{\d^2}{\d t^2}S(\rho(t))\right|_{t=0}$ is strictly negative unless $\rho=\rho_D$.

An alternative proof of eqn. (\ref{nort}), again using the inequality (\ref{with}), is as follows.       For an $N$ state system, there are $2^N$  matrices  that are 
diagonal matrices (in some chosen basis)  with diagonal matrix elements that are all $\pm 1$.  Let $U_i$ be any of these and set $\rho_i = U_i \rho U_i^{-1}$.   Of course,
$\rho_i$ is also a density matrix, since $U_i$ is unitary.     The average of the $\rho_i$, over
all $2^N$ choices of $U_i$,
is the diagonal density matrix $\rho_D$. So eqn. (\ref{with}) says that the average of $S(\rho_i)$ is less than or equal to $S(\rho_D)$.     But $S(\rho_i)$ is independent of $i$ and equal
to $S(\rho)$, since the von Neumann entropy is invariant under conjugation by a unitary matrix such as $U_i$.    So in fact the average of the $S(\rho_i)$ is just $S(\rho)$ and
  the inequality (\ref{with}) becomes $S(\rho)\leq S(\rho_D)$.

Somewhat similarly to what we have explained here, 
concavity of the function $f(q)=-q\log q$ could  have been used in the classical arguments in section \ref{classic}, though we circumvented this by using Stirling's
formula instead.

\subsection{Conditional and Relative Quantum Entropy}\label{crq}

It is now possible to formally imitate some of the other definitions that we made in the classical case.  For example, if $AB$ is a bipartite system,
we define what is called  {\bf quantum conditional entropy}
\begin{equation}\label{70} 
S(A|B)=S_{AB}-S_B.
 \end{equation}
This name is potentially misleading because there is not a good quantum notion of conditional probabilities.  
  Unlike the classical case, quantum conditional entropy  is not an entropy conditional on something.  Nevertheless, in section \ref{qco},
  we will discuss at least one sense in which quantum conditional entropy behaves in a way analogous to classical conditional entropy.

There is also a fundamental difference from the classical case:
 quantum mechanically, $S(A|B)$ can be negative.  In fact, suppose that system $AB $ is in an entangled pure state.  Then $S_{AB}=0$ but
as system $B$ is in a mixed state, $S_B>0$.   So in this situation $S(A|B)<0$.

Another classical definition that is worth imitating is the mutual information.   Given a bipartite system $AB$ with density matrix $\rho_{AB}$, the {\bf quantum mutual
information} is defined just as it is classically:
\begin{equation}\label{7one} 
I(A;B)=S_A-S_{AB}+S_B.  \end{equation}
Here, however, we are more fortunate, and the quantum mutual
information is nonnegative:
\begin{equation}\label{7two}  
I(A;B)\geq 0. \end{equation}
Moreover, $I(A;B)=0$ if and only if the density matrix factorizes, in
the sense that 
\begin{equation}\label{7three} 
\rho_{AB}=\rho_A\otimes \rho_B. 
\end{equation}
Positivity of mutual information is also called subadditivity of entropy.   To begin with, quantum mutual information is a formal definition and it is not obvious how it
is related to information that one can gain about system $A$ by observing system $B$.  
 We will explore at least one aspect of this question in section \ref{encoding}.

Before proving positivity of mutual information, I will explain an interesting corollary.
  Although conditional entropy 
$S(A|B)$ can be negative, the possibility of  ``purifying'' a density matrix  gives a lower bound on $S(A|B)$.  Let $C$ be such that $ABC$ is in a pure
state.   Remember that in general if $XY$ is in a pure state then $S_X=S_Y$.   So if $ABC$ is in a pure state then $S_{AB}=S_C$ and $S_B=S_{AC}$.
Thus
\begin{equation}\label{7four} 
S_{AB}-S_B=S_C-S_{AC}\geq -S_A, \end{equation}
where the last step is positivity of mutual information.  So
\begin{equation}\label{7five} 
S(A|B)=S_{AB}-S_B\geq -S_A. \end{equation}
Reversing the roles of $A$ and $B$ in the derivation, we get the Araki-Lieb inequality \cite{AL}
\begin{equation}\label{alin} S_{AB}\geq |S_A-S_B|. \end{equation}
It is saturated if $S_{AB}=0$, which implies $S_B=S_A$.   What has just been explained is a typical argument exploiting the existence
of purifications.

Just as in the classical case, to understand positivity of the mutual information, it helps to first define the  {\bf quantum  relative entropy} \cite{Umegaki}.   Suppose that $\rho$ and $\sigma$
are two density matrices on the same Hilbert space $\H$.  The relative
entropy can be defined by imitating the classical formula:
\begin{equation}\label{7six} 
 S(\rho||\sigma)=\Tr \rho (\log \rho -\log
\sigma).  \end{equation}
For now, this is just a formal definition, but we will learn in section \ref{hyptest} that 
$S(\rho||\sigma)$ has the same interpretation quantum mechanically
that it does classically: if one's hypothesis is that a quantum system is described by a density matrix $\sigma$, and it is
actually described by a different density matrix $\rho$, then to learn that one is wrong, one
needs to observe $N$ copies of the system where
$N S(\rho||\sigma)>>1. $

Just as classically, it turns out that $S(\rho||\sigma)\geq 0$ for all density matrices $\rho,\sigma$, with equality precisely if $\rho=\sigma$.  
To prove  this, first diagonalize $\sigma$.  In general $\rho$ is not diagonal in the same basis.
Let $\rho_D$ be the diagonal density matrix obtained from $\rho$ by dropping the off-diagonal matrix elements in the basis in which $\sigma $ is
diagonal, and keeping the diagonal ones.   Since $\Tr\,\rho\log\sigma=\Tr\,\rho_D\log \sigma$, it follows directly from the definitions of von Neumann
entropy and relative
entropy that
\be\label{morro} S(\rho||\sigma)=S(\rho_D||\sigma)  +S(\rho_D)-S(\rho). \ee
This actually exhibits $S(\rho||\sigma)$ as the sum of two nonnegative terms.  We showed in eqn. (\ref{nort}) that $S(\rho_D)-S(\rho)\geq 0$.
As for $S(\rho_D||\sigma)$, it is nonnegative, because if
$\sigma=\diag(q_1,\dots,q_n)$, $\rho_D=\diag(p_1,\dots, p_n)$, then
\begin{equation}\label{80} 
S(\rho_D||\sigma)=\sum_i p_i(\log p_i-\log q_i), \end{equation}
which can be interpreted as a classical relative entropy and so is nonnegative. 
To get equality in these statements, we need $\sigma=\rho_D$ and $\rho_D=\rho$, so $S(\rho||\sigma)$ vanishes only if $\rho=\sigma$.

Now we can use positivity of the relative entropy to prove that  $I(A;B)\geq 0$ for any density
matrix $\rho_{AB}$.   Imitating the classical proof, we define
\begin{equation}\label{7eight} 
\sigma_{AB}=\rho_A\otimes\rho_B, \end{equation}
and we observe that 
\begin{equation}\label{7nine} 
\log \sigma_{AB}=\log \rho_A\otimes 1_B+1_A\otimes \log
\rho_B, \end{equation}  
 so
\begin{align}\label{number4} S(\rho_{AB}||\sigma_{AB})&=\Tr_{AB} \rho_{AB}(\log \rho_{AB}-\log 
\sigma_{AB})\cr &=\Tr_{AB}\rho_{AB}(\log \rho_{AB}-\log \rho_A\otimes 1_B-1_B\otimes \log\rho_B) \cr &
=S_A+S_B-S_{AB}=I(A;B).\end{align}
So just as classically, positivity of the relative entropy implies positivity of the mutual information (which is also called subadditivity of entropy).

The inequality (\ref{with}) that expresses the concavity of the von Neumann entropy can be viewed as a special case of the positivity
of mutual information.  Let $B$ be a quantum system with density matrices $\rho_B^i$ and let $C$ be 
an auxiliary system $C$ with an orthonormal basis $|i\ra\neg_C$.
Endow $CB$ with the density matrix:
\begin{equation}\label{144}
   \rho_{CB}=\sum_i p_i|i\ra\neg_C\,_C\la i|\otimes \rho_B^i. \end{equation}
  The mutual information between $C$ and $B$ if the combined system is described
  by $\rho_{CB}$ is readily computed to be
\begin{equation}\label{145}
  I(C;B)= S(\rho_B)-\sum_i p_i S(\rho_B^i), \end{equation}
so positivity of mutual information gives our inequality.

\subsection{Monotonicity of Relative Entropy}

So relative entropy is positive, just as it is classically.      Do we dare to hope that relative entropy is also monotonic,
as classically?     Yes it is, as first proved by Lieb and Ruskai \cite{LR}, using a lemma of Lieb \cite{L}.
How to prove strong subadditivity will not be described here; this has been explored in a companion
article \cite{EW}, sections 3 and 4.

Monotonicity of quantum relative entropy is something of a miracle, because, as there is no such thing
as a joint probability distribution for general quantum observables, the intuition behind the classical statement is not applicable in any obvious way.  
Rather, strong subadditivity is ultimately used to prove that quantities such as quantum conditional entropy and quantum relative entropy and quantum mutual
information do have properties
somewhat similar to the classical case.   We will explore some of this in section \ref{more topics}.

There are different statements of monotonicity of relative entropy, but a very basic one (and actually the version proved in \cite{LR}) is monotonicity
under partial trace.  If $AB$ is a bipartite system with two density matrices $\rho_{AB}$ and $\sigma_{AB}$, then we can  take a partial trace on
$B$ to get reduced density matrices on $A$:
\begin{equation}\label{9two}
 \rho_A=\Tr_B\rho_{AB},~~~\sigma_A=\Tr_B\sigma_{AB}. \end{equation}
Monotonicity of relative entropy under partial trace is the statement
that taking a partial trace can only reduce the relative entropy:
\begin{equation}\label{9three}
S(\rho_{AB}||\sigma_{AB})\geq S(\rho_A||\sigma_A). \end{equation}
(This is also called the Data Processing Inequality.)

By imitating what we said classically in section \ref{classic}, one can deduce strong subadditivity of quantum entropy from monotonicity of relative entropy.
   We consider a tripartite system $ABC$
with density matrix $\rho_{ABC}$.  There are reduced density matrices such as $\rho_{A}=\Tr_{BC}\rho_{ABC}$, $\rho_{BC}=\Tr_{A}\rho_{ABC}$, etc., and we define
a second density matrix 
\begin{equation}\label{9four}
\sigma_{ABC}=\rho_{A}\otimes \rho_{BC}.   \end{equation}
The reduced density matrices of $\rho_{ABC}$ and $\sigma_{ABC}$,
obtained by tracing out $C$, are
\begin{equation}\label{9five}
\rho_{AB}=\Tr_C\rho_{ABC},~~~~~~~~\sigma_{AB}=\Tr_C\sigma_{ABC}=\rho_A\otimes \rho_B.  \end{equation}

Monotonicity of relative entropy under partial trace  says that
\begin{equation}\label{9six}
S(\rho_{ABC}||\sigma_{ABC})\geq
S(\rho_{AB}||\sigma_{AB}). \end{equation}
But (as in our discussion of positivity of mutual information)
\begin{equation}\label{9seven}
S(\rho_{ABC}||\sigma_{ABC})=S(\rho_{ABC}||\rho_A\otimes\rho_{BC})=
I(A;BC)=S_A+S_{BC}-S_{ABC} \end{equation}
and similarly
\begin{equation}\label{9eight}
S(\rho_{AB}||\sigma_{AB})=S(\rho_{AB}||\rho_A\otimes
\rho_B)=I(A;B)=S_A+S_B-S_{AB}. \end{equation}

So eqn. (\ref{9six})  becomes {\bf monotonicity of mutual information}
\begin{equation}\label{9nine}
I(A;BC)\geq I(A;B) \end{equation}
or equivalently {\bf strong subadditivity} \cite{LR}
\begin{equation}\label{100}
 S_{AB}+S_{BC}\geq S_B+S_{ABC}.  \end{equation}

All of these steps are the same as they were classically.
Using purifications, one can find various equivalent statements.  If $ABCD$ is in a pure state then $S_{AB}=S_{CD}$,  $S_{ABC}
=S_D$ so the inequality becomes
\begin{equation}\label{101}
S_{CD}+S_{BC}\geq  S_B+S_D. \end{equation}
So for instance $S(C|D)=S_{CD}-S_D$ can be negative, or
$S(C|B)=S_{BC}-S_B$ can be negative, but
\begin{equation}\label{102}
S(C|D)+S(C|B)\geq 0. \end{equation}
(This is related to {\bf monogamy of entanglement}: a given qubit in $C$ can be entangled with $D$, reducing $S_{CD}$,  or with $B$, reducing $S_{BC}$, but not both.)

Classically, the intuition behind monotonicity of mutual information was explained in section \ref{classic}; one learns at least as much about system $A$ by
observing $B$ and $C$ as one could learn by observing $B$ only.   Quantum mechanically, it is just not clear {\it a priori} that the formal definition 
$I(A;B)=S_A-S_{AB}+S_B$ will lead to something
consistent with that intuition.   The rather subtle result of monotonicity of relative entropy \cite{LR} shows that it does.

In general, strong subadditivity (or monotonicity of relative entropy) is the key to many  interesting statements in quantum information theory.
Many of the most useful statements that are not more elementary are deduced from strong subadditivity.

\subsection{Generalized Measurements}\label{gm}

Once we start using density matrices, there are a few more tools we should add to our toolkit.    First let us discuss measurements.
   Textbooks begin with ``projective measurements,'' which involve projection onto orthogonal subspaces of a Hilbert space $\H$ of quantum states. 
   We pick orthogonal  hermitian  projection operators $\pi_s$,
   $s=1,\cdots, k$ obeying
\begin{equation}\label{104}
\sum_s \pi_s=1,~~~~\pi_s^2=\pi_s,~~~~~~
\pi_s\pi_{s'}=0,~~s\not=s'.  \end{equation}
A measurement of a state $\psi$ involving these projection operators has outcome $s$ with
probability
\begin{equation}\label{105}
p_s=\la\psi|\pi_s|\psi\ra. \end{equation}
These satisfy $\sum_sp_s=1$ since $\sum_s\pi_s=1$.   If instead of a pure state $\psi$ the system is described by a density matrix $\rho$, then the 
probability of outcome $s$ is
\begin{equation}\label{106}
p_s=\Tr_\H \,\pi_s\rho. \end{equation}
After the measurement is made, if outcome $s$ has been found, the system can be described by a new density matrix
\be\label{belb}\rho_s=\frac{1}{p_s}\pi_s\rho\pi_s. \ee

But Alice can make a more general type of measurement using an auxiliary system $C$  (sometimes called an ancillary system) with Hilbert space $\CC$.   
We suppose that $\CC$ is $k$-dimensional with a basis of states $|s\ra$, $s=1,\cdots,k$.      Alice initializes $\CC$ in the  state $|1\ra$.   Then
she acts on the combined system $\CC\otimes \H$ with a unitary transformation $U$, which she achieves by suitably adjusting a time-dependent Hamiltonian.  
She chooses $U$ so that for any $\psi \in\H$
\begin{equation}\label{107}
U(|1\ra\otimes \psi)=\sum_{s=1}^k |s\ra \otimes E_s
\psi  \end{equation}
for some linear operators $E_s$.  (She does not care  what $U$ does on other states.)   
Unitarity of $U$ implies that
\begin{equation}\label{108}
\sum_{s=1}^k E_s^\dagger E_s = 1, \end{equation}
but otherwise the $E_s$ are completely arbitrary.  

Then Alice makes a projective measurement of the system $\CC\otimes \H$, using the commuting projection
operators
\begin{equation}\label{109}
\pi_s = |s\ra\la s| \otimes 1, \end{equation} 
which have all the appropriate properties.  The probability of outcome
$s$ is
\begin{equation}\label{110}
p_s=|E_s|\psi\ra|^2=\la\psi|E_s^\dagger
E_s|\psi\ra. \end{equation}
More generally, if the system $\H$ is described initially by a density matrix $\rho$, then the
probability of outcome $s$ is
\begin{equation}\label{111}
p_s=\Tr\,E_s^\dagger E_s\rho. \end{equation}
The numbers 
$p_s$ are nonnegative because $E_s^\dagger E_s$ is nonnegative, and   $\sum_s p_s=1$ because $\sum_s E_s^\dagger E_s=1$.      But the $E_s^\dagger E_s$ are not
orthogonal projection operators; they are just nonnegative hermitian operators that add to 1.     What we have described is a more general kind of quantum mechanical
measurement of the original system. (In the jargon, the positive operators $E_s^\dagger E_s$ whose sum is 1 comprise   a ``positive operator-valued measure'' or POVM.) 

According to eqn. (\ref{belb}), after  Alice's  measurement, if the outcome $s$ has been found, then the combined system $\CC\otimes \H$ can
be described by the density matrix $\frac{1}{p_s}|s\ra\la s| \otimes E_s|\psi\ra\la \psi|E_s^\dagger$.    Taking the trace over Alice's system,
the original system, after the measurement, can then be described by the density matrix 
 \be\label{zolbo}\frac{1}{p_s} E_s|\psi\ra\la \psi|E_s^\dagger, \ee
or more generally by $\frac{1}{p_s}E_s\rho E_s^\dagger$, if the original system was initially in a mixed state with density matrix $\rho$.
If after acting with $U$, Alice simply discards the subsystem $\CC$, or if this subsystem
is  inaccessible and we have no information about it,  then at that point the original system can be described by the density matrix
\be\label{opogo}\sum_s E_s|\psi\ra\la\psi|E_s^\dagger,\ee or more generally  by
$\sum_s E_s\rho E_s^\dagger$.

One can slightly generalize this construction as follows.\footnote{The following paragraph may be omitted on first reading.  It is included to make possible
a more general statement  in section \ref{qc}. }
Suppose that the initial system actually had for its Hilbert space a direct sum
$\H\oplus \H'$, but it is known that the initial state of the system is valued in $\H$, in other words the initial state $\psi$ has the form $\chi\oplus 0$
with $\chi\in\H$, and 0 the zero vector in $\H'$.   Then Alice couples $\H\oplus \H'$ to her auxiliary system $C$, so she describes the combined system
by a Hilbert space $\CC\otimes (\H\oplus \H')$.  Now she picks $U$ so that it maps a vector $|1\ra \otimes (\chi\oplus 0)$ to $\sum_s |s\ra\otimes (0\oplus E_s\chi)$,
where $E_s$ is a linear transformation $E_s:\H\to \H'$.   (As before, Alice does not care what $U$ does on other vectors.)   After applying $U$, Alice makes
a projective measurement using the same projection operators $\pi_s=|s\ra\la s|\otimes 1$ as before (of course, 1 is now the identity on $\H\oplus\H'$). The linear
transformations
$E_s$ still obey eqn. (\ref{108}),
the probability of outcome $s$ is still given by eqn. (\ref{111}), and the density matrix after a measurement that gives outcome $s$ is still given by eqn.
(\ref{zolbo}).

\subsection{Quantum Channels}\label{qc}

Now let us view this process from another point of view.     How can a density matrix evolve?    The usual Hamiltonian evolution of a state $\psi$
is $\psi\to U\psi$ for a unitary operator $U$, and on the density
matrix  it corresponds to 
\begin{equation}\label{112}
\rho\to U\rho U^{-1}. \end{equation}
As we remarked earlier (eqn. (\ref{zold})), such unitary evolution preserves the von Neumann entropy of a density matrix, and similarly
it preserves the relative entropy between two density matrices.

   But let us consider Alice again with her extended system
   $\CC\otimes \H$.    She initializes the extended system with the density matrix
\begin{equation}\label{113}
\h\rho =|1\ra\la1|\otimes \rho \end{equation}
where $\rho$ is a density matrix on $\H$.  Then she applies the same
unitary $U$ as before, mapping $\h\rho$ to
\begin{equation}\label{114}
\h\rho'= U\h\rho U^{-1}=\sum_{s,s'=1}^k |s\ra\la s'| \otimes E_s\rho
E_{s'}^\dagger. \end{equation}
 The induced density matrix on the original system $\H$ is obtained by
 a partial trace and is
\begin{equation}\label{115}
 \rho'=\Tr_\CC \h\rho'=\sum_{s=1}^k E_s\rho E_s^\dagger. \end{equation}

 We have found a more general way that density matrices can evolve.
 The operation
\begin{equation}\label{116}
 \rho \to \sum_{s=1}^k E_s \rho E_s^\dagger,~~~~~~ \sum_s
 E_s^\dagger E_s=1  \end{equation}
 is called a ``quantum channel,'' and the $E_s$ are called Kraus operators.  Unitary evolution is the special
 case in which there is only one Kraus operator.
 
 The notion of a quantum channel is axiomatized in more complete treatments than we will
 give here.\footnote{In the most general case, a quantum channel is a ``completely positive trace-preserving'' (CPTP) map
 from density matrices on one Hilbert space $\H$ to density matrices on another Hilbert space $\H'$.}    The upshot of a general analysis is that the most general
 physically sensible evolution of a density matrix  takes the form (\ref{116}), provided one allows the generalization described at the end of section  \ref{gm} in 
 which the $E_s$ are linear transformations from one Hilbert space $\H$ to another Hilbert space $\H'$.

 Now let $\rho$ and $\sigma$ be two different density matrices on $\H$.    Let us ask what happens to the relative entropy $S(\rho||\sigma)$ when we apply
 a quantum channel, mapping $\rho$ and $\sigma$ to
\begin{equation}\label{117}
 \rho'=\sum_s E_s\rho E_s^\dagger,~~~~~ \sigma'= \sum_s E_s\sigma
 E_s^\dagger. \end{equation}
 The first step of initialization, replacing $\rho$ and $\sigma$ by $|1\ra\la 1|\otimes \rho$ and $|1\ra\la 1|\otimes \sigma$, 
 does not change anything.  
    The second step,
 conjugating by a unitary matrix $U$, also does not change anything since  relative entropy is invariant under conjugation.    
 Finally, the last step was a partial trace, which can only reduce the quantum relative entropy.    So relative entropy can only go down under a quantum channel:
 $$S(\rho||\sigma)\geq S(\rho'||\sigma').$$     This is the most general statement of monotonicity of quantum relative entropy. 
 
We conclude this section with some exercises to familiarize oneself with quantum channels.

(1) Let $\psi$ be any pure state of a given system.  Find Kraus operators of a quantum channel that maps any density matrix $\rho$ to $|\psi\ra\la\psi|$.
(One way to implement this is to turn on a Hamiltonian for which $\psi$ is the ground state, and wait until the system relaxes to its ground state
by releasing energy to the environment.)
 
(2) Find Kraus operators of a quantum channel that maps any density matrix  for a given system (with finite-dimensional Hilbert space)
to a maximally mixed one, a multiple of the identity.   (This can arise as the outcome of sufficiently random interaction of the system with its environment.)

(3) Do the same for a quantum channel that, in a given basis, maps any $k\times k$ density  matrix 
$\rho=(\rho_{ij})$  to the corresponding diagonal density matrix
$\rho_D=\mathrm{diag}(\rho_{11},\rho_{22},\cdots,\rho_{kk}).$  (An idealized description of a physical realization is as follows.  A cavity is probed by atoms.
Denote as $|n\ra$ the state of the cavity when it contains $n$ photons.    Suppose that $n$ is unchanged when an atom
passes through the cavity, but the final state of the atom depends on $n$.  The probability to find the cavity in state $|n\ra$ is unchanged by the interaction with
a passing atom, so in the basis $\{|n\ra\}$, the diagonal elements of the density matrix are unchanged.   After many atoms have passed through the cavity,
an observation of the atoms would reveal with high confidence the number of photons in the cavity.  Therefore, tracing over the atomic states, the final
density matrix of the cavity is diagonal in the basis $\{|n\ra\}$.  Regardless of what state the cavity begins in, it will end up with high probability in an eigenstate
of the photon number operator, though one cannot say what the eigenvalue will be.)

(4) Show that the composition of two quantum channels is a quantum channel.  If the first channel has Kraus operators $E_s,$
$s=1,\cdots, p$, and the second has Kraus operators $E'_t$, $t=1,\cdots, q$, what are the Kraus operators of the composite
channel?

(5)  This and the next exercise involve  quantum channels  that  map one Hilbert space to another.  The goal is to show that natural operations that are
well-motivated in other ways can also be viewed as special cases of the evolution described in eqn. (\ref{116}).
First, given a Hilbert space $\H$,  construct a rather trivial quantum channel that maps density matrices on $\H$ to density matrices on a 1-dimensional
Hilbert space $\H_0$.   Note that, since a density matrix is hermitian, positive-definite, and of trace 1, there is a unique density matrix on $\H_0$,
namely the unit density  matrix 1.  Thus, given a Hilbert space $\H$, find Kraus operators $E_s:\H\to\H_0$ for a quantum channel that maps any density matrix $\rho$ on $\H$
to the density matrix 1 on $\H_0$.    Once you have done this, show that a partial trace is a quantum channel in the following sense.
If $AB$ is a bipartite system with Hilbert space $\H_A\otimes \H_B$, find Kraus operators $E_s:\H_A\otimes \H_B\to \H_A$ that implement the
partial trace $\rho_{AB}\to \rho_A=\Tr_B \rho_{AB}$.    In other words, find operators $E_s:\H_A\otimes \H_B \to \H_A$, satisfying $\sum_s E_s^\dagger E_s=1$
and $\sum_s E_s\rho_{AB} E_s^\dagger =\Tr_B \,\rho_{AB}$, for any $\rho_{AB}$.

(6) Let $A$ be a quantum system with Hilbert space $\H_A$, and let $B$ be a second quantum
system with Hilbert space $\H_B$ and some given density matrix $\rho_B$.   Find  Kraus operators
$E_s:\H_A\to\H_A\otimes \H_B$ for a quantum channel that combines a quantum system $A$ with some other system $B$ by 
mapping any given density matrix $\rho_A$ on $A$
to the density matrix $\rho_A\otimes \rho_B$ on $AB$. (You might want to consider first the trivial case that $\H_A$ is 1-dimensional.)
  An example of this is what happens whenever a system $A$ under study is combined
with some experimental apparatus $B$, which has been initialized in the state $\rho_B$.

 \subsection{Thermodynamics And  Quantum Channels}\label{exqu}
 
 As an example of these considerations, let us suppose that $\sigma$ is a thermal
 density matrix at some temperature $T=1/\beta$
\begin{equation}\label{118}
 \sigma=\frac{1}{Z}\exp(-\beta H).  \end{equation}
So $\log \sigma=-\beta H -\log Z$ and therefore the relative entropy between any density matrix $\rho$
 and $\sigma$ is
 \begin{align}\label{lok}S(\rho||\sigma)=&\Tr \,\rho(\log\rho -\log \sigma)=-S(\rho)+\Tr\rho( \beta H+\log Z) \cr=&\beta (E(\rho)-TS(\rho)) +\log Z\end{align}
 where the average energy computed in the density matrix $\rho$ is
\begin{equation}\label{119}
 E(\rho)=\Tr\,\rho H.   \end{equation}
We define the free energy 
\begin{equation}\label{120}
  F(\rho)=E(\rho)-T S(\rho). \end{equation}
The $\log Z$ term in eqn (\ref{lok}) is independent of $\rho$ and gives a
 constant that ensures that $S(\sigma||\sigma)=0$.    So
\begin{equation}\label{121}
 S(\rho||\sigma)=\beta (F(\rho)-F(\sigma)) . \end{equation}

   Now consider any evolution of the system, that is any quantum channel,
 that preserves thermal equilibrium at temperature $\beta$.     Thus, this channel maps $\sigma$ to itself, but it maps
 $\rho$ to a generally different density matrix $\rho'$. 
The relative entropy can only go down under a quantum channel, so
\begin{equation}\label{122}
 S(\rho||\sigma)\geq S(\rho'||\sigma),  \end{equation}  and therefore
\begin{equation}\label{123}
 F(\rho)\geq F(\rho').  \end{equation}

 In other words, a quantum channel that preserves thermal equilibrium can only reduce the free energy.
 This is an aspect of the second law of thermodynamics.   If you stir a system in a way that maps thermal equilibrium at temperature $T$ to
 thermal equilibrium at the same temperature, then it moves any density matrix closer to thermal equilibrium at temperature $T$.
 
To specialize further, take the temperature $T=\infty$, $\beta=0$.  (This makes sense for a system with a finite-dimensional Hilbert space.)  The thermal density matrix $\sigma$ is then maximally mixed, a multiple of the identity.
 For $T\to\infty$, $F(\rho)\sim  -TS(\rho)$.  So in this case, reducing the free energy means increasing the entropy.   Thus a quantum channel
 that maps a maximally mixed density matrix to itself can only increase the entropy.    The condition that a channel maps a maximally
 mixed density matrix to itself is $\sum_s E_s E_s^\dagger=1$.   (A channel satisfying this condition is called unital.   By contrast, the condition  $\sum_s E_s^\dagger E_s=1$  is satisfied
 by all quantum channels.)

An example of a quantum channel that maps a maximally mixed density matrix to itself is the channel
 that maps any density matrix $\rho$ to the corresponding diagonal density matrix $\rho_D$ (in some chosen basis).
The fact that the entropy can only increase under such a channel implies
the inequality $S(\rho)\leq S(\rho_D)$ (eqn. (\ref{nort})).

\section{More On Quantum Information Theory}\label{more topics}

From this point, one could pursue many different directions toward a deeper understanding of quantum information theory.  
This article will conclude with three topics that the author found helpful in gaining insight about the meaning of formal definitions such as quantum conditional
entropy and quantum relative entropy.   These concepts were defined by formally imitating the corresponding classical definitions, and it is not
really clear {\it a priori} what to expect  of such formal definitions. 

A secondary reason for the choice of topics is to help the reader appreciate the importance of monotonicity of quantum relative entropy -- and its
close cousin, strong subadditivity.  At several points, we will have to invoke monotonicity of relative entropy to prove that quantities like
quantum mutual information and quantum relative entropy that have been defined in a formal way do behave in a fashion suggested by their names.

The three topics that we will consider are quantum teleportation and conditional entropy, relative entropy and quantum hypothesis testing, and the use of a quantum state
to encode classical information.

\subsection{Quantum Teleportation and Conditional Entropy}\label{qco}

We start with {\bf quantum teleportation} \cite{bennett}.     For a first example,  imagine that Alice has in her possession a {\bf qubit}  $A_0$, a quantum
system with a two-dimensional Hilbert space.       Alice would like to help Bob create in his lab a qubit in a state identical to $A_0$.   
However, it is too difficult to actually send a qubit; she can only communicate by sending a classical message over the telephone.  
If Alice knows the state of her qubit, there is no problem: she tells Bob the state of her qubit and he creates one like it in his lab.    If, however, Alice
does not know the state of her qubit, she is out of luck.      All she can do is make a measurement, which will give some information about the prior
state of qubit $A_0$.     She can tell Bob what she learns, but the measurement will destroy the remaining information about $A_0$ and it will never
be possible for Bob to recreate $A_0$.

 Suppose, however, that Alice and Bob have previously shared a qubit pair $A_1$$B_1$ (Alice has $A_1$, Bob has $B_1$) in a known entangled
 state, for example
\begin{equation}\label{126}
 \Psi_{A_1B_1}=\frac{1}{\sqrt 2}\left(|0\,0\ra
   +|1\,1\ra\right)_{A_1B_1}.  \end{equation}
    Maybe Alice created this pair in her lab and then Bob took $B_1$ on the road with him, leaving $A_1$ in Alice's lab.  In this case, Alice can
 solve the problem.   To do so she makes a joint measurement of her system $A_0A_1$ in a basis that is chosen so that no matter what the answer is,
 Alice learns nothing about the prior state of $A_0$.   In the process, she also loses no information about $A_0$, since she had none before.   But as we will see, after getting her measurement outcome,
 she can  tell Bob what to do to recreate $A_0$.
 
 To see how this works, let us describe a specific measurement that Alice can make on $A_0A_1$ that will shed no light on the state of $A_0$.
    She can project $A_0A_1$ on the basis of four states
\begin{equation}\label{127}
 \frac{1}{\sqrt 2}(|0\,0\ra\pm |1\,1\ra )_{A_0A_1}~~
 \mathrm{and} ~~ 
 \frac{1}{\sqrt 2}(|0\,1\ra\pm |1\, 0\ra)_{A_0A_1} . \end{equation}
     To see the result of a measurement, suppose the unknown state of
     qubit $A_0$ is $\alpha|0\ra +\beta|1\ra$.    So the initial state
     of $A_0A_1B_1$ is
\begin{equation}\label{128}
 \Psi_{A_0A_1B_1}=\frac{1}{\sqrt 2}\left(\alpha |0\,0\,0\ra +\alpha
   |0\,1\,1\ra +\beta |1\,0\,0\ra +\beta
   |1\,1\,1\ra\right)_{A_0A_1B_1}.  \end{equation}

 Suppose that the outcome of Alice's measurement is to learn that $A_0A_1$ is in the state \be\frac{1}{\sqrt 2}(|0\,0\ra -|1\,1\ra)_{A_0A_1}.\ee   After the measurement,
  $B_1$ will be in the state
 $\left(
 \alpha|0\ra-\beta|1\ra\right)_{B_1}$.   Knowing this, Alice can tell Bob that he can recreate the initial state by acting on his qubit
 by 
\begin{equation}\label{129}
 \Psi_{B_1}\to \begin{pmatrix}1 & 0 \cr 0 &
   -1 \end{pmatrix}\Psi_{B_1} \end{equation} 
 in the basis $|0\ra$, $|1\ra$.    The other cases are similar, as the reader can verify.
 
 We will analyze a generalization, but first it is useful to formalize in a different way the idea that Alice is trying to teleport an arbitrary unknown quantum state.
    For this, we add another system $R$, to which Alice and Bob do not have access.     We assume that $R$ is maximally entangled with $A_0$ in a known
 state, say 
\begin{equation}\label{130}
  \Psi_{RA_0}=\frac{1}{\sqrt 2}\left(|0\,0\ra
    +|1\,1\ra\right)_{RA_0}. \end{equation}
    In this version of the problem, Alice's goal is to manipulate her system $A_0A_1$ in some way, and then tell Bob what to do to his system $B=B_1$ so that in the end the
 system $RB_1$ will be in the same state
\begin{equation}\label{131}
  \Psi_{RB_1}=\frac{1}{\sqrt 2}\left(|0\,0\ra
    +|1\,1\ra\right)_{RB_1} \end{equation}
  that $RA_0$ was previously -- with $R$ never being touched.    In this version of the problem, the combined system
  $RAB_1=RA_0A_1B_1$ starts in a pure state $\Psi_{RAB_1}=\Psi_{RA_0}\otimes \Psi_{A_1B_1}$.     The solution of this version of the problem is the same as the other one: Alice makes
  the same measurements and sends the same instructions as before. 
  
 We can understand better what is happening if we
  take a look at the {\it conditional entropy} of the system $AB=A_0A_1B_1$.     Since $A_1B_1$ is in a pure state, it does not contribute to $S_{AB}$,  so
  $S_{AB}=S_{A_0}=1$ ($A_0$ is maximally mixed, since it is maximally entangled with $R$).       Also $S_B=1$ since $B=B_1$ is maximally entangled with
  $A_1$.    Hence
\begin{equation}\label{132}
   S(A|B)=S_{AB}-S_B=1-1=0. \end{equation}
     It turns out that this is the key to quantum teleportation:
     teleportation, in a suitably generalized sense, is possible when and only when
\begin{equation}\label{133}
  S(A|B)\leq 0.  \end{equation}
  
  Let us explain first why this is a necessary condition.    We start with an arbitrary system $RAB$ in a pure state $\Psi_{RAB}$; Alice has access to $A$, Bob has access 
  to $B$, and neither one has access to $R$.      For teleportation, Alice might measure her system $A$ using some rank 1 orthogonal projection operators $\pi_i$.   (If she makes a more
  general measurement, for example using projection operators of higher rank, the system $RB$ does not end up in a known pure state and she will not be able to give appropriate
  instructions to Bob.)
  No matter what answer she gets, after the measurement, system $A$ is in a pure state and therefore $RB$ is also in a pure state $\chi_{RB}$, generally entangled.
     For teleportation, Alice has to choose the $\pi_i$ so that, no matter what outcome she gets, the density matrix $\rho_R$ of $R$ is the same as 
  before.    If this is so, then after her measurement, the state $\chi_{RB}$ of $RB$ is a purification of the original $\rho_R$.    
 Since she knows her measurement outcome, Alice knows which entangled state
is $\chi_{RB}$ and can convey this information to Bob.    Bob is then in possession of part $B$ of a known purification $\chi_{RB}$ of system $R$.   He makes in his lab
  a copy $A'$ of Alice's original system $A$, initialized in a known pure state $\Omega_{A'}$, so now he has part $A'B$ of a known purification $\widetilde \Psi_{RA'B}=\Omega_{A'}\otimes \chi_{RB}$ of $\rho_R$.
   By  a unitary transformation of system $A'B$, which Bob can implement in his lab, 
   $\widetilde\Psi_{RA'B}$ can be converted into any other pure state of $RA'B$ that purifies the same $\rho_R$.
   (This was explained following eqn. (\ref{Fourty9}).)   So Bob can convert  $\widetilde\Psi_{RA'B}$
   to a copy of the original $\Psi_{RAB}$.

 But do there exist projection operators of Alice's system with the necessary properties?    
The initial state $\Psi_{ABR}$ is pure so it has
\begin{equation}\label{134}
S_{AB}=S_R.   \end{equation}
    Bob's density matrix at the beginning is
\begin{equation}\label{135}
  \rho_B =\Tr_{RA}\,\rho_{RAB} \end{equation}
  where $\rho_{RAB}$ is the initial pure state density matrix.  
  By definition
\begin{equation}\label{136}
  S_B=S(\rho_B). \end{equation}
       If Alice gets measurement outcome $i$, then Bob's density matrix after the measurement
  is 
\begin{equation}\label{137}
  \rho_B^i=\frac{1}{p_i}\Tr_{RA}\,\pi_i \rho_{RAB}.  \end{equation} 

  Note that
\begin{equation}\label{138}
  \rho_B=\sum_i p_i\rho_B^i, \end{equation}
  since $\sum_i\pi_i=1$.    
  After the measurement, since $A$ is in a pure state, $RB$ is also in a pure state $\Psi^i_{RB}$, so $S(\rho_B^i)=S_R$.   But by hypothesis, the measurement
  did not change $\rho_R$, so  $S_R$ is unchanged and so equals the
  original $S_{AB}$.  Hence
\begin{equation}\label{139}
  S(\rho_B^i)=S_{AB}. \end{equation}    If all this is possible 
\begin{equation}\label{140}
  S_{AB}=S(\rho_B^i)=\sum_i p_i S(\rho_B^i). \end{equation}
The concavity inequality (\ref{with})  or equivalently positivity of the Holevo information (\ref{holevo}) says that if $\rho_B=\sum_i p_i\rho_B^i$ then
\begin{equation}\label{141}
  S(\rho_B)\geq \sum_i p_i S(\rho_B^i). \end{equation}
   So if
  teleportation can  occur, 
\begin{equation}\label{142}
  S_{AB}=\sum_i p_i S(\rho_B^i) \leq S(\rho_B)=S_B \end{equation}
  and hence $S(A|B)=S_{AB}-S_B\leq 0$.

Actually, $S(A|B)\leq 0$ is sufficient as well as necessary for teleportation, in the following sense       \cite{Horo}. (In this generality, what we are calling
teleportation is known as state merging.)
  One has to consider the problem of teleporting not a single system but $N$ copies of the system for large $N$.   (This is a common device
  in quantum information theory.  It is a rough  analog of the fact
  that to get simple statements in the classical case in section \ref{classic}, we had to consider a long message, obtained by sampling $N$ times from a probability
  distribution.)
     So one takes $N$ copies of system $RAB$ for large $N$, thus replacing $RAB$ by $R^{\otimes N} A^{\otimes N}B^{\otimes N}$.
      This multiplies all the entropies by $N$, so it preserves the condition $S(A|B)\leq 0$.      Now Alice tries to achieve teleportation by
  making a complete projective measurement on her system $A^{\otimes N}$.    It is very hard to find an explicit set of projection operators $\pi_i$ with the
  right properties, but it turns out, remarkably, that for large $N$, a random choice will work (in the sense that with a probability approaching 1, the error
  in state merging is vanishing for $N\to\infty$).     This statement actually has strong subadditivity as a corollary 
  \cite{Horo}.  This approach to strong subadditivity has been described in sections 10.8-9 of \cite{Preskill}.

We actually can now give a good explanation of the meaning of quantum conditional entropy $S(A|B)$.    Remember that classically $S(A|B)$ measures  
 how many additional bits of information Alice has to send to Bob after he has already received $B$, so that he will have full knowledge of $A$.  
We will find a quantum analog of this, but now involving qubits rather than classical bits.  Suppose that $S(A|B)>0$ and Alice nevertheless wants to share her state with Bob.
Now we have to assume that Alice is capable of quantum communication, that is of sending a quantum system to Bob while maintaining
its quantum state, but that she wishes to minimize the amount of quantum communication she will need.
  She first creates some maximally entangled qubit pairs and sends half of each pair to Bob.
 Each time she sends Bob half of a pair, $S_{AB}$ is unchanged but $S_B$ goes up by 1,  so $S(A|B)=S_{AB}-S_B$ goes down by 1.     So $S(A|B)$, if positive, is the
 number of such qubits that Alice must send to Bob to make $S(A|B)$ nonpositive and so make teleportation or state merging
  possible without any further quantum communication.

 If $S(A|B)$ is negative, teleportation or state merging 
 is possible to begin with and  $-S(A|B)$ is the number of maximally entangled qubit pairs that Alice and Bob can be left with afterwards \cite{Horo}.    This may be seen as follows.
 Alice creates an auxiliary system $A'A''$, where $A'$ consists of $n$ qubits that are completely entangled with another set of $n$ qubits that comprise system $A''$.   Alice
 considers the problem of teleporting to Bob the combined system $\bar A=A''A$, while leaving $A'$ untouched.   Since
  $S(\bar A|B)=n+S(A|B)$, Alice observes that $S(\bar A|B)<0$ provided $n<-S(A|B)$.   Given this
 inequality, 
 Alice can teleport $\bar A=A''A$ to Bob, keeping $A'$ in reserve.   At the end of this, Alice and Bob share $n$ maximally entangled qubit pairs, namely Alice's system $A'$ and Bob's copy of $A''$.
 This description is a shorthand; it is implicit that at each stage, we are free to replace the system under consideration by the tensor product of $N$ copies of itself, for some large $N$.
 As a result, integrality of $n$ is not an important constraint.   A more precise statement of the conclusion is that for large $N$, after teleportation to Bob of 
part $A^{\zotimes}$ of a composite system $A^\zotimes B^\zotimes$, Alice and Bob can be left with up to $-N S(A|B)$ maximally entangled qubit pairs.

 \subsection{Quantum Relative Entropy And Hypothesis Testing}\label{hyptest}
 
In a somewhat similar way, we can give a physical meaning to the relative entropy $S(\rho||\sigma)$ between two density matrices $\rho$, $\sigma$.
 Recall  from section \ref{cre} that classically, if we believe a random variable is governed by a probability distribution $Q$ but it is actually governed by a probability distribution $P$,
 then after $N$ trials the ability to disprove the wrong hypothesis is
 controlled by
\begin{equation}\label{146}
  2^{-NS(P||Q)}. \end{equation}
  
 A similar statement holds quantum mechanically:   if our initial hypothesis is that a quantum system $X$ has density matrix $\sigma$, and the actual answer is $\rho$,
 then after $N$ trials with an optimal measurement used to test the
 initial hypothesis, the confidence that the initial hypothesis was
 wrong is controlled in the same sense by
\begin{equation}\label{147}
 2^{-NS(\rho||\sigma)}. \end{equation}

Let us first see that monotonicity of relative entropy implies that one cannot do better than that \cite{HP}.     A measurement is a special case of a quantum channel,
 in the following sense.    To measure a system $X$, one lets it interact quantum mechanically with some other system $YC$ where $Y$ is any quantum system and $C$
 is the measuring device.       After they interact, one looks at the measuring device and forgets the rest.    Forgetting the rest is a partial trace that maps 
 a density matrix $\beta_{XYC}$ to $\beta_C=\Tr_{XY}\beta_{XYC}$.     If $C$ is a good measuring device with $n$ distinguishable quantum states, 
 this means that in a distinguished basis $|\alpha\ra$,
 $\alpha=1,\cdots, n$,
 its density matrix $\beta_C$ will have a diagonal form
\begin{equation}\label{148}
 \beta_C=\sum_\alpha b_\alpha|\alpha\ra\la\alpha|. \end{equation}
 The ``measurement'' converts the original density matrix  into the probability distribution $\{b_\alpha\}$.
 
 So when we try to distinguish $\rho$ from $\sigma$, we use a quantum channel plus partial trace
 (or simply a quantum channel, since a partial trace can be viewed as a quantum channel)
 that maps $\rho$ and $\sigma$ 
 into density matrices for $C$
\begin{equation}\label{149}
 \rho_C= \sum_\alpha r_\alpha
 |\alpha\ra\la\alpha|~~~~~\sigma_C=\sum_\alpha s_\alpha
 |\alpha\ra\la\alpha|, \end{equation}
and thereby into classical probability distributions $R=\{r_\alpha\}$ and $S=\{s_\alpha\}$.   We can learn that $\rho$ and $\sigma$ are different
 is by observing that $R$ and $S$ are different, a process controlled
 by
\begin{equation}\label{150}
 2^{-NS_{\mathrm{cl}}(R||S)}, \end{equation}
where  $S_{\mathrm{cl}}(R||S)$ is the classical relative entropy
 between $R$ and $S$.    
 
  This is the same as the  relative entropy
 between
$\rho_C$ and $\sigma_C$:
\begin{equation}\label{151}
 S(\rho_C||\sigma_C)=S_{\mathrm{cl}}(R||S). \end{equation}
And monotonicity of relative entropy gives
\begin{equation}\label{152}
 S(\rho||\sigma)\geq S(\rho_C||\sigma_C). \end{equation} 
 So if we follow this procedure, then $S(\rho||\sigma)$ gives a bound on how well we can
 do:
\begin{equation}\label{153}
 2^{-NS_{\mathrm{cl}}(R||S)}\geq 2^{-N S(\rho||\sigma)}. \end{equation}
 
 Actually, quantum mechanics allows us to do something more sophisticated than making $N$ repeated measurements of the system of interest.    We could more
 generally make a joint measurement on all $N$ copies. 
 Taking $N$ copies replaces the Hilbert space $\H$ of the system under study by $\H^\zotimes$, and replaces the density matrices $\sigma$ and $\rho$
 by $\sigma^\zotimes$ and $\rho^\zotimes$. All entropies and relative entropies are multiplied by $N$.
 A joint measurement on $N$ copies would convert a density matrix $\sigma^\zotimes$ or $\rho^\zotimes$ to a probability distribution
 $S^{[N]}$ or $R^{[N]}$.     We will not learn much from a single joint measurement on $N$ copies, since it will just produce a random answer.   But given $NN'$ copies
 of the system, we could repeat $N'$ times a joint measurement of $N$ copies.   The ability to distinguish $S^{[N]}$ from $R^{[N]}$ in $N'$ tries is controlled for large $N'$ by
 $2^{-N' S_{\mathrm{cl}}(R^{[N]}||S^{[N]})}$. The monotonicity of relative entropy gives 
$2^{-N'S_{\mathrm{cl}}(R^{[N]}||S^{[N]})}\geq 2^{-N'  S(\rho^\zotimes||\sigma^\zotimes)}=2^{-\widehat N S(\rho||\sigma)}$, where $\widehat N=NN'$.
   So also with such a more general procedure, the ability to disprove in $\widehat N$ trials  an initial hypothesis $\sigma$ for a system actually described by $\rho$
   is bounded by $2^{-\widehat N S(\rho||\sigma)}$.
 
In the limit of large $\widehat N$, it is actually possible to saturate this
bound, as follows \cite{Hayashi, Hayashi2}.
 If $\rho$ is diagonal in the same basis in which $\sigma$ is diagonal, then by making a measurement that involves projecting on 1-dimensional
 eigenspaces of $\sigma$, we could convert the density matrices $\rho$, $\sigma$ into classical probability distributions $R,S$ with
 $S(\rho||\sigma)=S_{\mathrm{cl}}(R||S)$.  The quantum problem would be equivalent to a classical problem,  even without taking many copies.   As usual the subtlety comes because the matrices
 are not simultaneously diagonal.  By dropping from $\rho$ the off-diagonal matrix elements in some basis in which $\sigma$ is diagonal,
 we can always construct a diagonal density matrix $\rho_D$.  Then a measurement projecting on 1-dimensional eigenspaces of $\sigma$
 will give probability distributions $R,S$ satisfying
\begin{equation}\label{154}
 S(\rho_D||\sigma)=S_\cl(R||S).\end{equation}
 This is not very useful, because it is hard to compare $S(\rho_D||\sigma)$ to $S(\rho||\sigma)$.  That is why it is necessary to consider a joint measurement on $N$ copies, for large $N$,
  which  makes possible
an easier alternative to comparing $S(\rho_D||\sigma)$ to $S(\rho||\sigma)$, as we will
 see.
 
  Let us
 recall the definition of relative entropy:
\begin{equation}\label{155}
 S(\rho^\zotimes||\sigma^\zotimes)=\Tr\,\rho^\zotimes \log
 \rho^\zotimes - \Tr\,\rho^\zotimes
 \log\sigma^\zotimes. \end{equation} 
 The second term $\Tr\,\rho^\zotimes \log\sigma^\zotimes$ is unchanged
 if we replace $ \rho^\zotimes$ by its counterpart $(\rho^\zotimes)_D$ that is diagonal
 in the same basis as $\sigma^\zotimes$.    So
\begin{equation}\label{156}
 S(\rho^\zotimes||\sigma^\zotimes)-S((\rho^\zotimes)_D||\sigma^\zotimes)=\Tr
 \rho^\zotimes \log \rho^\zotimes
 -\Tr(\rho^\zotimes)_D\log(\rho^\zotimes)_D. \end{equation}   Actually,  there are
 many bases in which $\sigma^\zotimes $ is diagonal; it will be important to choose the right one in defining $(\rho^\zotimes)_D$.
 For large $N$, and with the right choice of basis, we will be able to get a useful bound on the right hand side of eqn. (\ref{156}).

Roughly speaking, there is simplification for large $N$ because group theory can be used to simultaneously put
 $\rho^\zotimes$ and $\sigma^\zotimes$ in a block diagonal form with relatively small blocks.  This will make possible the comparison we need.
 In more detail, the group $S_N$ of permutations of $N$ objects acts in an obvious way on $\H^\zotimes$.   It commutes with the action on $\H^\zotimes$
 of $U(k)$, the group of unitary transformations of the $k$-dimensional Hilbert space $\H$.   Schur-Weyl duality gives the decomposition of 
 $\H^\zotimes$ in irreducible representations of $S_N\times U(k)$.   Every Young diagram ${\D}$ with $N$ boxes and at most $k$ rows
determines an  irreducible representation $\lambda_{\D}$ of $S_N$ and an irreducible representation $\mu_{\D}$ of $U(k)$.   The decomposition
of $\H^\zotimes$ in irreducibles of $S_N\times U(k)$ is 
 \be\label{werro}\H^\zotimes =\oplus_{\D} \lambda_{\D}\otimes \mu_{\D}.\ee
 The $\lambda_{\D}$ of distinct ${\D}$ are non-isomorphic, and the same is true of the $\mu_{\D}$.   Let $a_{\D}$ and $b_{\D}$ be, respectively, the dimension
 of $\lambda_{\D}$ and of $\mu_{\D}$.  The maximum value of $b_{\D}$ is bounded\footnote{See eqn. (6.16) of \cite{Hayashi2}.   One approach
 to this upper bound is as follows.   In general, the highest weight of an irreducible representation of the group $SU(k)$ is a linear combination of certain
 fundamental weights with nonnegative integer coefficients $a_i$, $i=1,\cdots,k-1$.  In the case of a representation associated to a Young diagram with $N$ boxes,
 the $a_i$ are bounded by $N$.   The dimension of an irreducible representation with highest weights $(a_1,a_2,\cdots,a_{k-1})$ is a polynomial in the $a_i$
 of total degree $k(k-1)/2$, so if all $a_i$ are bounded by $N$, the dimension is bounded by a constant times $N^{k(k-1)/2}$.   One way to prove that the 
 dimension is a polynomial in the $a_i$ of the stated degree is to use the Borel-Weil-Bott theorem.  According to this theorem, a representation with highest
 weights $(a_1,a_2,\cdots,a_{k-1})$ can be realized as $H^0(F,\otimes_{i=1}^{k-1} \L_i^{a_i})$, where 
 $F=SU(k)/U(1)^{k-1}$ is the flag manifold  of the group $SU(k)$ and $\L_i\to F$ are certain holomorphic line bundles.
 Because $F$ has complex dimension $k(k-1)/2$, the Riemann-Roch theorem says that the dimension of $H^0(F,\otimes_{i=1}^{k-1} \L_i^{a_i})$ is
 a polynomial in the $a_i$ of that degree.}
  by a power of $N$:
 \be\label{ferro}b_{\mathrm{max}}\leq (N+1)^{k(k-1)/2}.\ee
 The important point will be that $b_{\mathrm{max}}$ grows only polynomially for $N\to\infty$, not exponentially.  In contrast,
 the numbers $a_\D$ can be exponentially large for large $N$.
 
 Eqn. (\ref{werro}) gives a decomposition of $\H^\zotimes$ as the direct sum of subspaces of dimension $a_{\D}b_{\D}$.   Since $\rho^\zotimes$ and $\sigma^\zotimes$
 commute with $S_N$, they are block diagonal with respect to this decomposition.   But more specifically, the fact that $\rho^\zotimes$ and $\sigma^\zotimes$
 commute with $S_N$ means that each $a_{\D}b_{\D}\times a_{\D}b_{\D}$ block is just the direct sum of $a_{\D}$ identical blocks of size $b_{\D}\times b_{\D}$.
  So $\rho^\zotimes$ has a decomposition
 \begin{equation}\label{157}
 \rho^\zotimes=\begin{pmatrix} p_1\rho_1 &&& \cr & p_2\rho_2 &&\cr
   && p_3\rho_3 & \cr &&&\ddots \end{pmatrix} \end{equation}
   in blocks of size $b_{\D}\otimes b_{\D}$, with each such block occurring $a_{\D}$ times, for all possible ${\D}$.  (The total number of blocks is
   $\sum_\D a_\D$.)       The $\rho_i$ are density matrices and the $p_i$ are nonnegative numbers adding to 1. In the same basis, $\sigma^\zotimes$ has just the same sort of
   decomposition:
  \begin{equation}\label{157a}
 \sigma^\zotimes=\begin{pmatrix} q_1\sigma_1 &&& \cr & q_2\sigma_2 &&\cr
   && q_3\sigma_3 & \cr &&&\ddots \end{pmatrix} .\end{equation}
 We can furthermore make a unitary transformation in each block to diagonalize $\sigma^\zotimes$.    This will generically not diagonalize
 $\rho^\zotimes$.   But because $\rho^\zotimes $ is block diagonal with relatively small blocks, its entropy can be usefully 
  compared with that of the diagonal density matrix $(\rho^\zotimes)_D$ that is obtained by setting to 
 0 the off-diagonal matrix elements of $\rho^\zotimes$ in a basis in which $\sigma^\zotimes$ is diagonal within each block and keeping the diagonal ones:
 \begin{equation}\label{158}
  (\rho^\zotimes)_D=\begin{pmatrix} p_1\rho_{1,D} &&& \cr &
    p_2\rho_{2,D} &&\cr && p_3\rho_{3,D} & \cr
    &&&\ddots \end{pmatrix} .\end{equation}

 One finds then 
\begin{equation}\label{159}
 \Tr \rho^\zotimes \log \rho^\zotimes
-\Tr(\rho^\zotimes)_D\log(\rho^\zotimes_D)=\sum_i p_i
(S(\rho_{iD})-S(\rho_i)).\end{equation}
It is important that a potentially large term $\sum_i p_i \log p_i$ cancels out here.
Any density matrix on an $n$-dimensional space has an entropy $S$ bounded by $0\leq S \leq \log n$.   Because the sizes of the blocks are bounded above by 
$b_{\mathrm{max}}\sim N^{k(k-1)/2}$, and $\sum_i p_i=1$, the right hand side\footnote{The right hand side is actually positive because
of the inequality (\ref{nort}).} of eqn. (\ref{159}) is bounded  by  $\log b_{\mathrm{max}}\sim\frac{1}{2}k (k-1)\log N$, 
which for large $N$ is negligible compared to $N$.

\def\zut{{[N]}}
Combining this with eqns. (\ref{154}) and (\ref{156}), we see that 
for large $N$, a measurement that projects onto 1-dimensional eigenspaces of $\sigma_i$ within each block
maps the density matrices $\rho^\zotimes$ and $\sigma^\zotimes$ to classical probability distributions $R^\zut$ and $S^\zut$ such that the quantum
relative entropy $S(\rho^\zotimes||\sigma^\zotimes)$ and the classical relative entropy $S(R^\zut||S^\zut)$ are asymptotically equal. To be more precise,
$S(\rho^\zotimes||\sigma^\zotimes)=NS(\rho||\sigma)$ is of order $N$ for large $N$, and differs from $S(R^\zut||S^\zut)$ by at most a constant times $\log N$.
In other words
\be\label{kino}S(\rho||\sigma)=\frac{1}{N} S(\rho^\zotimes||\sigma^{\zotimes})=\frac{1}{N}S(R^\zut||S^\zut)+\mathcal{O}\left(\frac{\log N}{N}\right). \ee

Once we have identified a measurement that converts the quantum relative entropy (for $N$ copies of the original system) to a classical relative entropy,
we take many copies again and invoke the analysis of classical relative entropy in section \ref{cre}.   In more detail,
  consider a composite system
consisting of $N$ copies of the original system.   Suppose that we observe $N'$ copies of this composite system (making $NN'$ copies of the original system), 
for very large $N'$.  On each
copy of the composite system, we make the above-described measurement.  This means that we sample $N'$ times from the classical probability distribution
$S^\zut$ (if the original hypothesis $\sigma$ was correct) or $R^\zut$ (if the original system was actually described by $\rho$).  According to the classical
analysis in section \ref{cre}, the ability to distinguish between $R^\zut$ and $S^\zut$ in $N'$ trials is controlled by $2^{-N'S(R^\zut||S^\zut)}$.   
According to eqn. (\ref{kino}), this is asymptotically the same as $2^{-N' S(\rho^\zotimes||\sigma^\zotimes)}=2^{-NN' S(\rho||\sigma)}$.
In short, we learn that after a suitable measurement on $\widehat N=NN'$ copies of the original system, we can distinguish between the hypotheses $\sigma$ and $\rho$
with a power
\be\label{yeft} 2^{-\widehat N  S(\rho||\sigma)},\ee
saturating the upper bound (\ref{153}) (with the total number of trials now being $\widehat N$ rather than $N$).  In the exponent, there are errors of order $N'\log N$ (from the logarithmic
correction in (\ref{kino})) and $N\log N'$ (coming from the fact that  the classical analysis of section \ref{cre}, which for instance used only the leading
term in Stirling's formula, has corrections of relative order $\frac{1}{N'}\log N'$).

This confirms that quantum relative entropy has the same interpretation as classical relative
 entropy: it controls the ability to show, by a measurement, that an initial hypothesis is incorrect. 
A noteworthy fact \cite{Hayashi} is that the measurement that must be made on the composite system to accomplish this depends only on $\sigma$ (the initial
hypothesis) and not on $\rho$ (the unknown answer).
 
At the outset, we assumed monotonicity of relative entropy and deduced from it an upper bound (\ref{147}) on how well one can distinguish two density matrices
 in $N$ trials.  Actually, now that we know that the upper bound is attainable, one can reverse the argument and show that this upper bound implies
 monotonicity of relative entropy.  Suppose that $AB$ is a bipartite system with density matrices $\rho_{AB}$, $\sigma_{AB}$ that we want to distinguish
 by a measurement.  One thing that we can do is to forget system $B$ and just make measurements on $A$.  The above argument shows that, after taking
 $N$ copies, the reduced density matrices $\rho_A=\Tr_B\,\rho_{AB}$, $\sigma_A=\Tr_B\,\sigma_{AB}$ can be distinguished at the rate $2^{-NS(\rho_A||\sigma_A)}$.
 But since measurements of subsystem $A$ are a special case of measurements of $AB$, this implies that $\rho_{AB}$ and $\sigma_{AB}$ can be distinguished
 at the rate $2^{-NS(\rho_A||\sigma_A)}$.    If therefore we know the bound (\ref{147}), which says that $\rho_{AB}$ and $\sigma_{AB}$ cannot be distinguished at a faster
 rather than $2^{-N S(\rho_{AB}||\sigma_{AB})}$, then the monotonicity inequality $S(\rho_{AB}||\sigma_{AB})\geq S(\rho_A||\sigma_A)$ follows.  
In \cite{BSS}, monotonicity of relative entropy has been proved by giving an independent proof of the upper bound on how well two density matrices
can be distinguished.

 \subsection{Encoding Classical Information In A Quantum State}\label{encoding}

Finally, we will address the following question:   how many bits of information can Alice send to Bob by sending him a quantum system $X$ with a
 $k$-dimensional Hilbert space $\H$?    (See \cite{Preskill}, especially section 10.6, for more on this and related topics.)
 
 One thing Alice can do is to send one of $k$ orthogonal basis vectors in $\H$.     Bob can find which one she sent by making a measurement.    So in that way
 Alice can send $\log k$ bits of information.      We will see that in fact it is not possible to do better.
 
We suppose that Alice wants to encode a random variable  that takes the values $x_i$, $i=1,\dots,n$ with probability $p_i$.  When the value is $x_i$,
 she writes down this fact in her notebook $C$ 
   and creates a density matrix $\rho_X^i$ on system $X$.   If $|i\ra$ is the state of the notebook when Alice has written the value $x_i$, then on the combined
   system $CX$, 
Alice has created the density matrix
\begin{equation}\label{161}
 \rho_{CX} =\sum_i p_i |i\ra\la i| \otimes \rho_X^i \end{equation}
    Then Alice sends the system $X$ to Bob.   Bob's task is to somehow extract information by making a measurement.
 
Before worrying about what Bob can do, let us observe that the density matrix $\rho_{CX}$ of the system $CX$
 is the one (eqn. (\ref{144})) that was used earlier in discussing the entropy inequality for mixing. It is sometimes called a classical-quantum density matrix.
 The reduced density matrix of $X$ is $\rho_X=\Tr_C\,\rho_{CX}=\sum_i p_i \rho_X^i$.
  As before, the
 mutual information between $C$ and $X$ is the Holevo information \be\label{hdf}I(C;X)= S(\rho_X) -\sum_i p_i S(\rho_X^i).\ee
 Since $S(\rho_X^i)\geq 0$ and $S(\rho_X)\leq \log k$, it follows that
\begin{equation}\label{162}
  I(C;X)\leq \log k.   \end{equation}
 If we knew that quantum mutual information has a similar interpretation to classical mutual information, we would stop here and say that
 since $I(C;X)\leq \log k$, at most $\log k$ bits of information about the contents of Alice's notebook have been encoded in $X$.     However,
 we aim to demonstrate that quantum mutual information behaves like classical mutual information, at least in this respect, not to assume it.  As we will
 see, what we want is precisely what monotonicity of mutual information says, in the present context.

What can Bob do on receiving system $X$?  The best he can do is to combine it with some other system which may include a quantum system $Y$ and
a measuring apparatus $C'$.  He acts on the combined system $XYC'$ with some unitary transformation or more general quantum channel and then reads $C'$. 
The combined operation is a quantum channel.
 As in our discussion of relative
entropy, the outcome of the channel is a density matrix of the form 
\begin{equation}\label{163}
\rho_{C'}=\sum_{\alpha =1}^r q_\alpha
|\alpha\ra\la\alpha|, \end{equation}
where $|\alpha\ra$ are distinguished states of $C'$ -- the states that one reads in a classical sense.  
The outcome of Bob's measurement is a probability distribution $\{q_\alpha\}$ for a random variable  whose
values are labeled by $\alpha$.    What Bob learns about the contents of Alice's notebook is the classical mutual information between Alice's probability
distribution $\{p_i\}$ and Bob's probability distribution $\{q_\alpha\}$.   Differently put, what Bob learns is the mutual information $I(C;C')$.

To analyze this, we note that before Bob does anything, $I(C;X)$ is the same as $I(C;XYC')$ because $YC'$ (Bob's auxiliary quantum system $Y$
and his measuring apparatus $C'$) is not coupled to $CX$.     In more detail, the initial description of the combined system $CXYC'$ is  by the tensor product
of a density matrix $\rho_{CX}$  for $CX$ and a density matrix $\rho_{YC'}$
for $YC'$.   As one can deduce immediately from the definitions, the mutual information between $C$ and $XYC'$ if the full system $CXYC'$
is described by $\rho_{CX}\otimes \rho_{YC'}$ is the same as the mutual information between $C$ and $X$ if the subsystem $CX$ is described by
$\rho_{CX}$.
Bob then acts on $XYC'$ with a unitary transformation, or maybe a more general quantum channel, which can only reduce the mutual information.
Then he takes a partial trace over $XY$, which also can only
reduce the mutual information, since monotonicity of mutual information under partial trace
tells us that
\begin{equation}\label{164}
I(C;XYC')\geq I(C;C'). \end{equation}  
So 
\begin{equation}\label{165}
\log k\geq I(C;X) =I(C;XYC')_{\mathrm{before}}\geq
I(C;XYC')_{\mathrm{after}} \geq
I(C;C')_{\mathrm{after}}, \end{equation}
where ``before'' and ``after'' mean before and after Bob's manipulations.   More briefly, any way that Bob processes the signal he receives
can only reduce the mutual information.    Thus Alice cannot encode more than $\log k$ bits of classical information
in an $k$-dimensional quantum state, though it takes strong subadditivity (or its equivalents) to prove this.

The problem that we have discussed also has a more symmetrical variant.
In this version,   Alice and Bob share a bipartite state $AB$; Alice has access to $A$ and Bob
has access to $B$.   The system is initially described by a density matrix $\rho_{AB}$.  Alice makes a generalized measurement of $A$ and Bob makes
a generalized measurement of $B$.  What is the maximum amount of information that Alice's results may give her about Bob's measurements, and vice-versa?
An upper bound is given by the mutual information $I(A;B)$ in the initial density matrix $\rho_{AB}$.   Alice's measurements amount to a quantum channel
mapping her system $A$ to her measurement apparatus $C$; Bob's measurements amount to a quantum channel mapping his system $B$ to his measurement
apparatus $C'$.   The mutual information between their measurement outcomes is simply the mutual information $I(C;C')$ in the final state.   Monotonicity of mutual
information in any quantum channel says that this can only be less than the initial $I(A;B)$.

A more subtle issue is the extent to which these upper bounds can be saturated.  For an introduction to such questions
see \cite{Preskill}, section 10.6.

\vskip1cm
Research supported in part by NSF Grant PHY-1606531.    I thank N. Arkani-Hamed, J. Cotler, B. Czech, M. Headrick, and R. Witten  for discussions.  I also thank M. Hayashi, as well as the
referees,
for some explanations and helpful criticisms and for a careful reading of the manuscript.
\bibliographystyle{unsrt}

\end{document}